\documentclass[11pt]{article}

\usepackage{geometry}             
\usepackage{amsmath, amssymb, amsthm, amsfonts}
\usepackage{mathtools} 

\usepackage{graphicx}
\usepackage[numbers]{natbib}

\usepackage[ruled,vlined]{algorithm2e}
\usepackage{algorithmic}

\usepackage{hyperref}
\usepackage{xcolor}
\usepackage{booktabs} 
\usepackage{multirow}

\usepackage{tikz}
\usepackage{enumitem}

\theoremstyle{definition}
\newtheorem{Definition}{Definition}

\newlength\fheight \newlength\fwidth
\definecolor{mycolor1}{rgb}{1.00000,0.00000,1.00000}%

\newcommand{\specialcell}[2][c]{%
	\begin{tabular}[#1]{@{}c@{}}#2\end{tabular}}

\newcommand{\subcap}[1]{{\footnotesize #1}}

\title{Variational Multiscale Nonparametric Regression: Algorithms and Implementation}
\author{Miguel del Alamo$^{1,\dagger}$, Housen Li$^{2,\dagger}$, Axel Munk$^{2,}$\footnote{Correspondence: munk@math.uni-goettingen.de.} and Frank Werner$^{3,}$\footnote{These authors contributed equally to this work.}}

\date{%
$^{1}$ Department of Applied Mathematics,
University of Twente\\
$^{2}$ Institute for Mathematical Stochastics, University of Göttingen\\
$^{3}$ Institute of Mathematics, University of Würzburg}

\begin{document}
\maketitle

\begin{abstract}
Many modern statistically efficient methods come with tremendous computational challenges, often leading to large-scale optimisation problems. In this work, we examine such computational issues for recently developed estimation methods in nonparametric regression with a specific view on image denoising. We consider in particular certain variational multiscale estimators which are statistically optimal in minimax sense, yet computationally intensive. Such an estimator is computed as the minimiser of a smoothness functional (e.g., TV norm) over the class of all estimators such   that none of its coefficients with respect to a given multiscale dictionary is statistically significant. The so obtained multiscale Nemirowski--Dantzig estimator (MIND) can incorporate any convex smoothness functional and combine it with a proper dictionary including wavelets, curvelets and shearlets. The computation of MIND in general requires to solve a high-dimensional constrained convex optimisation problem with a specific structure of the constraints induced by the statistical multiscale testing criterion. To solve this explicitly, we discuss three different algorithmic approaches: the Chambolle--Pock, ADMM and semismooth Newton algorithms. Algorithmic details and an explicit implementation is presented and the solutions are then compared numerically in a simulation study and on various test images. We thereby recommend the Chambolle--Pock algorithm in most cases for its fast convergence. We stress that our analysis can also   be transferred to signal recovery and other denoising problems to recover more general objects whenever it is possible to borrow statistical strength from data patches of similar object structure.
\end{abstract}



\section{Introduction}

Regression analysis has a centuries long history in science and is one of the most powerful and   widely used tools in modern statistics. As it aims to discover a functional dependency $f$ between certain variables of interest, it provides important insight into the relationship of such variables. Typically, the data are noisy and specific regression models provide a mathematical framework for the recovery of this unknown relationship $f$ given such noisy observations. Due to the flexibility of such models, they can be accommodated to many different scenarios, e.g., for linearly dependent data as in linear regression and for data following a general nonlinear structure as in nonlinear regression. The literature on this and the range of applications is vast, we exemplarily refer to the work of \citet{DrSm98} for an overview. Often, these models depend on a few parameters only specifying $f$ already, which then can be estimated from the data in a relatively simple way. In contrast, nonparametric regression avoids such a restrictive modelling and becomes indispensable when prior knowledge is not sufficient for parametric modelling (see,   e.g., \cite{BoAz97,fan1996local}). One of the most fundamental and most studied nonparametric models is the Gaussian nonparametric regression (i.e.,    denoising) model with independent errors, sometimes denoted as white noise regression model (after a Fourier transformation). In this model, we aim to estimate the unknown regression function $f:\,[0,1]^d\rightarrow \mathbb{R}$ given noisy observations $Y_i$, which are related to $f$ as
\begin{equation}
Y_i=f(x_i)+\sigma\, \epsilon_i, \ \ x_i\in\Gamma_n, \ \ i=1,\ldots,n,\label{model}
\end{equation}
where $\epsilon_i$ are independent normal random variables with zero mean and unit variance    and   $\Gamma_n$ is a discrete grid in $[0,1]^d$ that consists of $n$ points. Note that $\sigma$ determines the noise level, which we assume for simplicity to be constant (see, however, Section \ref{CD}).  We stress that much of what is addressed in this paper can be generalised to other error models and other domains than the $d$-dimensional unit cube $[0,1]^d$ (see also Section \ref{CD}). However, to keep the presentation simple, we restrict ourselves to the Gaussian error and equidistant grids in the $d$-dimensional unit cube. A mathematical theory of such nonparametric regression problems has a long history in statistics (see \cite{stone1982optimal} for an early reference), as they are among the simplest models where the unknown object is still in a complex function space and not just encoded in a low-dimensional parameter, yet they are general enough to cover many applications (see,   e.g., \cite{fan1996local}). Consequently, the statistical estimation theory of nonparametric regression is a well investigated area for which plenty of methods have been proposed, such as kernel smoothing \cite{nadaraya1964estimating, watson1964smooth}; global regularisation techniques such as penalised maximum likelihood \cite{eggermont2000maximum}; ridge regression, which amounts to Tikhonov regularisation \cite{phillips1962technique, morozov1966regularization}; or total variation (TV) regularisation \cite{rudin1992nonlinear}. These methods have not been designed a-priori in a spatially adaptive way which could be overcome by a second generation of (sparse) localising regularisation methods originating in the development of wavelets \cite{daubechies1992ten}. To fine tune the estimator for first generation methods, usually a simple regularisation parameter has to be chosen (statistically); for wavelet-based estimators, this amounts to properly select (and truncate) the wavelet coefficients methods, e.g., by soft thresholding (see, e.g., \cite{donoho1995noising}). We refer to \citet{tsybakov2008introduction} for an introduction to the modern statistical theory of nonparametric regression, mainly from a minimax perspective. One of the latest developments to the problem of recovering the function $f$ in (\ref{model}) already dates back to Nemirovski \cite{nemirovskii1985nonparametric} and is implicitly exemplified by the Dantzig selector \cite{candes2007dantzig},  which  can be seen as a hybrid between a sparse approximation with respect to a dictionary and variational ($L^1$) regularisation. Such hybrid methods   were coined by \citet{grasmair2015} as MIND estimators (MultIscale Nemirovski--Dantzig) and are the focus of this paper. We   discuss these mainly in the context of statistical image analysis ($d=2$), but stress that our findings also apply to signal recover $d=1$ and to other situations where multiscale approaches are advantageous, e.g., for temporal-spatial imaging, where $d=3,4$. 

\subsection{Variational Denoising}

One of the most prominent regularisation methods for image analysis is total variation (TV) denoising, which was popularised by Rudin, Osher and Fatemi \cite{rudin1992nonlinear} (for further developments towards a mathematical theory for more general variational methods, see,   e.g., \cite{scherzer2009variational} and   references  therein). In the spirit of Tikhonov regularisation, the rationale behind is to enforce certain properties for the function $f$ fitted to the model \eqref{model}, which are encoded by a convex penalty term $R$ such as the TV seminorm. Consequently, the weighted sum of a least-squares data fidelity term (corresponding to the maximum likelihood estimation in model \eqref{model}), $R$ is minimised over all functions $g$ (e.g., of bounded variation)    and   the minimiser is taken as the reconstruction or estimator for $f$, i.e.,
\begin{equation}\label{eq:var}
\hat f \in \arg\min_g \left[ \frac12 \sum_{i=1}^n \left(Y_i - g \left(x_i\right)\right)^2 + \alpha R \left(g\right) \right]
\end{equation}
with a weighting factor (sometimes called the regularisation parameter) $\alpha > 0$. As soon as $R$ is convex, algorithmically, this leads to a convex optimisation problem, which can be solved efficiently in practice. This also applies to other convex data fidelity, e.g., when the likelihood comes from an exponential family model. To solve (\ref{eq:var}) numerically, most common are first-order methods  (see,   e.g., \cite{bss16}), which boil down to computing the (sub-)gradients of the least-squares term and $R$. 

One of the statistical disadvantages of methods of the form \eqref{eq:var} is the usage of the global least-squares term, which to some extend enforces the same smoothness of $g$ everywhere on the domain $\Gamma_n$. To overcome this issue, a spatially varying choice of the parameter $\alpha$ has been discussed in case of $R$ being the TV seminorm  (see,   e.g., \cite{hrc14,hpr17}). Other options are localised least-squares fidelities  \cite{hlrw18,dhrc11a,dhrc11b}  or anisotropic total variation penalties, where the weighting matrix $A$ is also improved iteratively based on the available data (see,   e.g., \cite{lb15}). We   discuss another powerful strategy to cope with local inhomogeneity of the signal which has its origin in statistics and has been recently shown to be statistically optimal (in a certain sense to be made precise below) for various choices of penalties $R$. 

\subsection{Statistical Multiscale Methods}
In the nonparametric statistics literature, multiscale methods have their origin in the discovery of wavelets \cite{daubechies1992ten}, which Donoho and Johnstone \cite{donoho1994ideal} used to construct wavelet thresholding estimators and showed their ability to adapt to certain smoothness classes. Such estimators are computationally simple, as they only involve thresholding with respect to an orthonormal basis (cf. \cite{donoho1995noising}). Following wavelets,    myriad   multiscale dictionaries tailored to different needs have been developed, such as curvelets \cite{candes2000curvelets}, shearlets \cite{labate2005sparse}, etc. In a nutshell, the superiority of multiscale dictionaries over other bases for denoising and inverse problems lies in their excellent approximation and localising properties, which yields sparse approximations of functions with respect   to loss functions that results from norms which average the error, e.g., $L^p$, $p\geq1$, Sobolev or Besov norms. However, it is well known that despite this sparseness these estimators tend to present Gibbs-like oscillatory artefacts in statistical settings (which is not well reflected by such norms). This affects the quality of the overall reconstruction critically \cite{CandesGuo, del2018frame}. 

\subsection{Variational Multiscale Methods}

Variational multiscale methods offer a solution to the problem of such Gibbs-artefacts, as they combine multiscale methods with classical (non-sparse) regularisation techniques, with the idea of making use of the best of both worlds: the stringent data-fitting properties of (overcomplete) multiscale dictionaries,  and   the desired (problem dependent) smoothness imposed by a regularisation method, e.g., by TV regularisation.

\begin{Definition}
Let $\mathcal{F}$ be a class of functions, $R:\mathcal{F}\to \mathbb{R}\cup\{\infty\}$ a convex functional    and   $\{\phi_{\lambda}| \lambda\in\Lambda_n\}$ a finite collection of functions. Given observations $Y_i$ as in \eqref{model}, the MultIscale Nemirovski--Dantzig Estimator (MIND) is defined by the constrained optimisation problem
\begin{equation}
\hat{f}\in \underset{g\in\mathcal{F}}{\textup{argmin}}\ R(g)\ \ \textup{ such   that } \max_{\lambda\in\Lambda_n} |\langle \phi_{\lambda},g-Y\rangle|\leq q_n\label{MIND}
\end{equation}
for a threshold $q_n>0$, where we use the inner product $\langle h,g\rangle:=n^{-1}\sum_{i=1}^nh(x_i)g(x_i)$.
\end{Definition}

Several particular cases of the MIND have been proposed: the first time by Nemirovski \cite{nemirovskii1985nonparametric} (who gave credit to S. V. Shil'man) in 1985 for a system of indicator functions and a Sobolev seminorm, $R(g)=|g|_{H^s}$, Donoho \cite{donoho1995noising} derived soft-thresholding from this principle 
where the dictionary $\{\phi_{\lambda}\, |\, \lambda\in\Lambda_n\}$ is a wavelet basis  (for the latter, see also \cite{malgouyres2002}). For the case of a curvelet frame, we refer to \cite{CandesGuo}   and   for a system of indicator functions and the Poisson likelihood to \cite{fmm12b,fmm13}, where $R(g)=|g|_{BV}$ is the TV seminorm.
Notice that, in practice, the choices of $R$ and the dictionary reflect previous knowledge or expectations on the unknown function $f$: the amount of regularisation that $R$ imposes should be dependent on how smooth we expect $f$ to be; and the dictionary should measure patterns that we expect $f$ to obey, e.g., elongated features in the case of curvelets and  piecewise constant areas in the case of indicator functions. Furthermore, we have the following rule of thumb: the more redundant is the dictionary, the better is the statistical properties of the estimator but the more expensive is the computation of a solution to \eqref{MIND}. 

\begin{figure}
 \centering
 \begin{tabular}{cc}
 \includegraphics[width=0.27\textwidth]{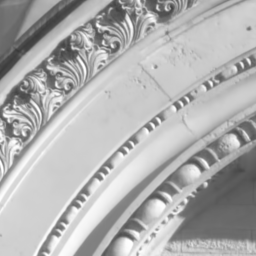} 
 &
 \includegraphics[width=0.27\textwidth]{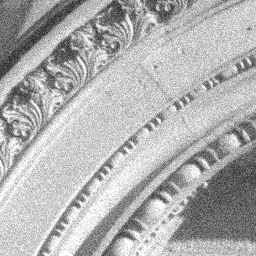} 
 \\
 \subcap{(\textbf{a}) Truth} & \subcap{(\textbf{b}) Noisy image, PSNR = 23.5} \\
 &
 \end{tabular}
 \begin{tabular}{ccc}
 \includegraphics[width=0.27\textwidth]{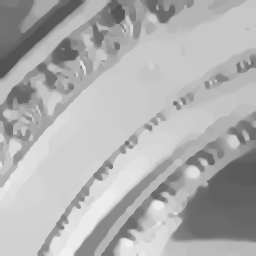} &
 \includegraphics[width=0.27\textwidth]{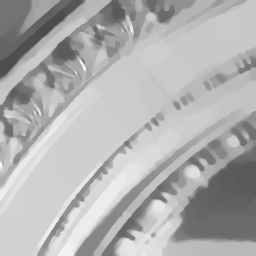} &
 \includegraphics[width=0.27\textwidth]{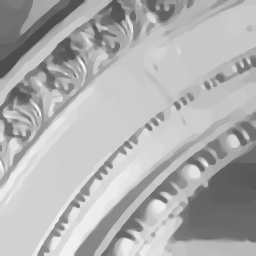} 
 \\
 \subcap{(\textbf{c}) MIND, wavelets, PSNR = 25.1} & \subcap{(\textbf{d}) MIND, curvelets, PSNR = 25.6} &\subcap{(\textbf{e}) MIND, shearlets, PSNR = 27.9} \\
 &&\\
 \includegraphics[width=0.27\textwidth]{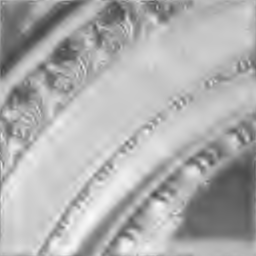}
 &
 \includegraphics[width=0.27\textwidth]{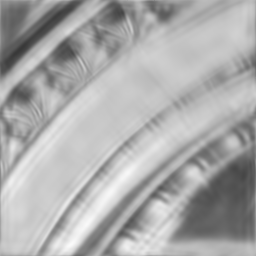} 
 & 
 \includegraphics[width=0.27\textwidth]{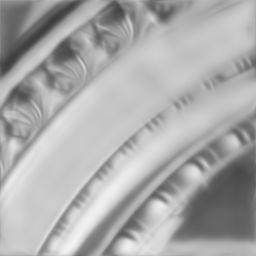} 
 \\
 \subcap{(\textbf{f}) ST, wavelets, PSNR = 24} & \subcap{(\textbf{g}) ST, curvelets, PSNR = 23.7} & \subcap{(\textbf{h}) ST, shearlets, PSNR = 24.4} 
 \end{tabular}
 \caption{Comparison of MIND with TV regularisation and the soft-thresholding (ST) estimator on the ``building'' image from Darmstadt Noise Dataset \cite{plotz2017benchmarking} with different dictionaries. Details of dictionaries can be found in Section \ref{ss:dict}. The thresholds for MIND and ST are chosen the same as the 50\% quantile of $\|\hat\sigma\epsilon\|_{\text{MS}}$ in \eqref{Residual}  (see Section \ref{ss:quant}). The noise level $\sigma$ is estimated by a second order difference-based estimator \cite{munk2005difference}  (see Section \ref{ss:nlvl}).}\label{f:intro}
\end{figure}

It is documented that the different proposed instances of the MIND show increased regularity and reduced artefacts as compared with standard multiscale methods (i.e., thresholding type estimators), but still avoid oversmoothing if the threshold $q_n$ is appropriately chosen  (see \cite{fmm12b,fmm13, del2018frame}). For instance, we illustrate in Figure \ref{f:intro} that the MIND with $R$ the TV seminorm outperforms the soft-thresholding estimator under various choices of multiscale dictionaries, including wavelets, curvelets and shearlets. In addition to their good empirical performance, estimators of the form \eqref{MIND} have recently been analysed theoretically and proved to be (nearly minimax) optimal for nonparametric regression \cite{grasmair2015,del2018frame} and certain inverse problems \cite{candes2007dantzig,del2019total}. We   give a brief review of these theoretical results in Section \ref{sec:theory}.

\subsection{Computational Challenges and Scope of the Paper}

However, the practical and theoretical superiority of regularised multiscale methods over purely dictionary thresholding methods comes at a cost: instead of simple thresholding, a complex optimisation problem has to be solved, with an increase in computation time and the need of improved optimisation methods. In practice, the problem \eqref{MIND} is non-smooth (e.g., if $R$ is chosen as the TV seminorm)   and   is furthermore high-dimensional due to a huge number of constraints. For instance, in the case of images of $256 \times 256$ pixels, the number of constraints is 1,751,915 for the dictionary of cubes of edge length $\le 30$ and 3,211,264 for that of shearlets. 

Besides our theoretical review in Section \ref{sec:theory}, the goal of this paper is to discuss different particularly successful algorithmic approaches for the solution of \eqref{MIND} in Section \ref{sec:comp}. Among the first attempts is the usage of the alternating direction method of multipliers (ADMM) in \cite{fmm12a,fmm12b,fmm13}, including a problem-specific convergence analysis. The computational disadvantage of this approach is the necessity to compute projections to the constraint set, which is most generally done by Dykstra's algorithm. Even though this can be efficiently implemented using GPUs (cf. \cite{lkhk15,khkl16}), the projection step provides a computational bottleneck of the overall algorithm. Several other convex optimisation algorithms such as Douglas--Rachford--Splitting (see \cite{m17}) or the Proximal Alternating Predictor Corrector algorithm (cf. \cite{ls18}) also require the computation of such projections,   and   hence their computational performance is similar to the ADMM-based version. In view of this, we   furthermore discuss two new approaches based on the Chambolle--Pock algorithm \cite{chambolle} on the one hand and a semismooth Newton method (cf. \cite{hintermueller,clason2018total}) on the other hand. The former allows   avoiding  computing projections to the constraint sets, but requires only the computation of the corresponding resolvent operator, which reduces to a high-dimensional soft-thresholding problem. The latter allows   solving  a regularised version of the optimality conditions by Newton's method and then apply a path-continuation scheme to decrease the amount of regularisation. Both algorithms hence avoid Dykstra's projection step, but still have favourable convergence properties. 

In Section \ref{sec:num}, we   finally discuss the practical advantages and disadvantages of the previously described algorithms along different numerical examples. For all examples checked in this paper, we found that the Chambolle--Pock algorithm is superior compared to the others. 
Finally, we provide a Matlab implementation of the Chambolle--Pock algorithm for this problem and code to run all examples in \url{https://github.com/housenli/MIND}.

\section{Theoretical Properties of Variational Multiscale Estimation Methods}\label{sec:theory}

In this section, we briefly review some theoretical reconstruction properties of multiscale variational estimators within model \eqref{model}. Recall that, in the nonparametric regression model \eqref{model}, we have access to noisy samples of a function $f$ at locations $x_i\in \Gamma_n$. We have some flexibility in choosing the underlying grid $\Gamma_n$: it could be for instance an equidistant $d$-dimensional grid (e.g., as pixels in an image), but other choices are possible (e.g., a polar grid).

\subsection{Theoretical Guarantees}

Estimators of the form \eqref{MIND} have been analysed from a theoretical viewpoint in a variety of settings. When the dictionary basis functions $\phi_{\lambda}$ are orthogonal, this becomes particularly simple, as then the evaluation functionals (if the truth equals $g$) 
$\langle \phi_{\lambda}, \epsilon \rangle$ become independent in model \eqref{MIND}. This is valid for wavelet systems and their statistical analysis is vast  (see,   e.g., \cite{Lep91,DoEt95,DoJo95,WeWa98,HKPT98,Cai02,Zhan05,ABDJ06,CaZh09} for various forms of adaptation and thresholding techniques). If the dictionary is redundant the analysis becomes more difficult (see, however,   \cite{haltmeier2014} for the asymptotic validity of hard and soft thresholding in this case) and only in recent years it could be been shown that suitably constructed MINDs with redundant dictionaries perform optimal in a statistical (minimax) sense over certain function spaces. {In the following,  we summarise some of these results in an informal way. To this end, the notion of minimax optimality is key, which compares a given estimator with the \textit{best} possible estimator in terms of their \textit{worst case error} (see Equation \eqref{minimax}    and   Definition 3.1 in \cite{tsybakov2008introduction} for a formal definition).}

\begin{enumerate}[leftmargin=2.2em,labelsep=5.5mm]
\item[-] \emph{Sobolev spaces}: The authors of  \cite{nemirovskii1985nonparametric,grasmair2015} analysed the MIND with $R(g)=|g|_{H^s}$ and $\{\phi_{\lambda}\}$ being a set of indicator functions of rectangles at different locations and scales. They showed that, for the choice $q_n=C\, \sigma\, \sqrt{\log n/n}$ for an explicit constant $C>0$, the MIND is minimax optimal up to logarithmic factors for estimating functions in the Sobolev space $H^s$. This means that the MIND's expected reconstruction error is of the same order as the error of the best possible estimator, i.e.,
\begin{equation}
1\leq \frac{\sup_{|f|_{H^s}\leq L} \mathbb{E}\|f-\hat{f}_{MIND}\|_{L^p}}{\textup{inf}_{\text{all estimators } \hat f} \sup_{|f|_{H^s}\leq L} \mathbb{E}\|f-\hat{f}\|_{L^p}}\leq C\, Polylog(n).\label{minimax}
\end{equation}
Besides minimax optimality, \citet{grasmair2015} also showed that the MIND with $H^s$ regularisation is also optimal for estimating functions in other Sobolev spaces $H^t$ for certain smoothness indices $t$, a phenomenon known as adaptation.
\item[-] \emph{Bounded variation}: In \cite{del2018frame}, the MIND with bounded variation regularisation $R(g)=|g|_{BV}$ was considered. It was shown that it is optimal in a minimax sense up to logarithms for estimating functions of bounded variation if $d=2$. For $d\geq 3$, the discretisation matters further and this could only be shown in a Gaussian white noise model. Such results hold for a variety of dictionaries $\{\phi_{\lambda}\}$, such as wavelet bases, mixed wavelet and curvelet dictionaries,   as well as    suitable systems of indicator functions of rectangles. 
\end{enumerate}

In addition to theoretical guarantees, these results also provide a way of choosing the threshold parameter $q_n$ in \eqref{MIND}. Indeed, both for Sobolev and for bounded variation regularisation, it is shown (see again \cite{grasmair2015,del2018frame}) that the choice $q_n=C\, \sigma\, \sqrt{(\log n)/n}$ for an explicit constant $C>0$ yields asymptotically optimal performance. The constant $C$ depends on the dimension $d$ and smoothness $s$ of the functions, on whether we consider Sobolev or BV regularisation,   and   on the dictionary $\{\phi_{\lambda}\}$ we employ.

\subsection{Practical choice of the threshold}\label{ss:quant}

Besides the theoretically (asymptotically) optimal choice of the parameter $q_n$, a Monte Carlo method for a finite sample choice was proposed by \citet{grasmair2015}. It is based on the observation that the multiscale constraint in \eqref{MIND} can be interpreted as a test statistic for testing whether the data $Y$ is compatible with the function $g$, in the sense of \eqref{model}. In fact, the "multiscales" come from not only performing one test, but many tests that focus on different features of $g$ of various sizes, locations and orientations. 
From this viewpoint, $q_n$ is interpreted as a critical value for a statistical test,   and   statistical testing theory suggests that $q_n$ should be chosen as a high quantile of the random variable
\begin{equation}
\|\sigma\epsilon\|_{\text{MS}}:=\sigma \max_{\lambda\in\Lambda_n}|n^{-1}\sum_{i=1}^n\,\epsilon_i\phi_{\lambda}(x_i)|.\label{Residual}
\end{equation}

This interpretation yields a practical way of choosing $q_n$:  we simply pre-estimate $\sigma$, simulate independent realisations of the noise $\epsilon$, compute their values in \eqref{Residual}    and   finally set $q_n$ to be a quantile of that sample. 
This choice of $q_n$ yields good practical performance (see Section \ref{sec:num}) and is compatible with the theoretically optimal choice for $n$ large enough. Methods for pre-estimating $\sigma$ from the data can be found, e.g., in \cite{munk2005difference}  {(see Section \ref{ss:nlvl} for simulations of the MIND estimator with estimated noise level).}

\section{Computational Methods}\label{sec:comp}

In this section, we   discuss different algorithmic approaches to compute the MIND $\hat f$ in \eqref{MIND}. First,  we stress that the optimisation problem \eqref{MIND} is a non-smooth, convex, high-dimensional optimisation problem, which allows in principle to exploit any optimisation method designed for such situations. In the following, we restrict to three different approaches that we found particularly  suited for our scenario. To set the notation in this section, we rewrite \eqref{MIND} as
\begin{equation}
\min_{v\in \mathbb{R}^n} J \left(v\right), \qquad J(v) := F(Kv)+G(v),\label{primal}
\end{equation}
where $F$ and $G$ are lower semi-continuous, proper convex functionals given by
\begin{align*}
&F(w):=1_{\leq 0}(w-KY-q_n)+1_{\leq 0}(-w+KY-q_n) \ \textup{ for } w\in\mathbb{R}^{\#\Lambda_n}, \\
&G(v):=R(v) \ \textup{ for } v\in\mathbb{R}^{n}.
\end{align*}
and $K$ is the linear (bounded) operator that maps from $\mathbb{R}^n$ to $\mathbb{R}^{\#\Lambda_n}$ and is defined by
\begin{equation*}
[Kg_n]_{\lambda}:=\langle \phi_{\lambda},g_n\rangle \ \ \textup{ for any } \lambda\in\Lambda_n.
\end{equation*}

Here, and in what follows, $1_{\leq 0}$ denotes the indicator function of the negative half-space,  which is
\[
1_{\leq 0} \left(v\right) = \begin{cases} 0 & v_i \leq 0 \text{ for all possible } i,\\ \infty & \text{else}. \end{cases}
\]

Due to the convexity of $F$ and $G$, the problem \eqref{primal} can equivalently be solved by finding a root of the subdifferential $\partial J(v)$ (see,   e.g., \cite{rockafellar2015convex}), which is a generalisation of the classical derivative. Such roots correspond to stationary points of the evolution equation
\[
\partial v\left(t\right) \in - \partial J \left(v(t)\right), \qquad t > 0.
\]

Two fundamental algorithms for the solution of this equation arise from applying the explicit or implicit Euler method, which leads to the steepest decent procedure
\begin{subequations}\label{eq:Euler}
\begin{equation}\label{eq:steepest_decent}
v_{k+1} \in \left(I - \lambda_k \partial J \right) \left(v_k\right), \qquad k \in \mathbb N
\end{equation}
and the so-called proximal point method
\begin{equation}\label{eq:proximal_point}
v_{k+1} \in \left(I + \lambda_k \partial J \right)^{-1} \left(v_k\right), \qquad k \in \mathbb N
\end{equation}
\end{subequations}
respectively, where $I$ denotes the identity operator and $\lambda_k > 0$ is a step size parameter. {For \eqref{eq:steepest_decent}, $\lambda_k$ should be chosen such that $\left(I - \lambda_k \partial J \right)$ is a contraction to ensure convergence. For continuously differentiable $J$,  it therefore suffices to choose $\lambda_k < L^{-1}$ with the Lipschitz constant $L$ of $\nabla J$;  in practice, one often uses $\lambda_k = \left(2\hat L\right)^{-1}$ with an estimator $\hat L$ of the Lipschitz constant (generated,  e.g., by a power method). In the case of \eqref{eq:proximal_point}, convergence can be obtained for any choice of $\lambda_k > 0$, but the computation of the so-called resolvent operator $\left(I + \lambda_k \partial J \right)^{-1}$} is clearly as difficult as the original problem. Motivated by {these two} methods, different algorithms have been suggested. Therefore, the subdifferential $\partial J(v)$ of $J$ is split into the subdifferentials of $F$ and $G$, which allow for a much simpler computation of the corresponding resolvent operators. This makes use of the formula
\begin{equation}\label{eq:subdiff}
\partial J(v) = K^* \partial F \left(Kv\right) + \partial G(v),
\end{equation}
which holds true whenever there exists some vector $v \in \mathbb R^n$ such that both $F$ and $G$ are finite and continuous at $Kv$ and $v$ respectively (see e.g., Prop. 5.6 in \cite{et76}). In our situation, this is the case whenever $R$ is continuous at a point in the interior of the feasible set. With the help of \eqref{eq:subdiff}, we can also derive the necessary and (due to convexity) sufficient first-order optimality conditions
\begin{subequations}\label{eq:opt}
\begin{align}
-K^*w &\in \partial G \left(v\right), \\
K v & \in \partial F^\star \left(w\right)\label{eq:opt2}
\end{align}
\end{subequations}
with the conjugate functional
\[
F^\star \left(w\right) := \sup_{v \in \mathbb R^n} \left[ w^\top v - F \left(v\right)\right].
\]

In the following, we present three different algorithms to solve \eqref{primal} either via a specific splitting of $\partial J$ or via the first-order optimality conditions: the Chambolle--Pock primal dual algorithm, an ADMM method    and   a semismooth Newton method.

\subsection{The Chambolle--Pock Method}\label{sec:CP}

The primal dual algorithm by Chambolle and Pock \cite{chambolle} is based on a reformulation of the optimality conditions \eqref{eq:opt} as fixed point equations. If we multiply the second condition by a parameter $\tau > 0$ and add $w$ on both sides yields
\[
w + \tau K v \in w + \tau \partial F^\star(w).
\]

Similarly, the first condition yields
\[
v - \delta K^*w \in v + \delta \partial G (v).
\]

Hence, the solutions $v$ and $w$ of the optimality conditions can be found by repetitively applying the resolvent operators of $G$ and the dual of $F$, which are given by
\begin{subequations}\label{eq:prox}
\begin{align}
\big(I+\tau\, \partial G\big)^{-1}(v)&:=\textup{argmin}_{x\in \mathbb{R}^n}\, \frac{\|x-v\|^2}{2\tau}+R(v)
\\
\big(I+\delta\, \partial F^{\star}\big)^{-1}(w)&:=\textup{argmin}_{z\in \mathbb{R}^{\#\Lambda_n}}\, \frac{\|z-w\|^2}{2\delta} + q_n\sum_{\lambda\in\Lambda_n}\big|z_{\lambda}-\delta Y_{\lambda}\big|.\label{proxF}
\end{align}
\end{subequations}

This combined with an extrapolation step yields the Chambolle--Pock algorithm. It can also be interpreted as a splitting of the subdifferential $\partial J$ into those of $F$ and $G$,   and   then applying proximal point steps (i.e., backwards steps) to $\partial G$ and the dual of $\partial F$.

The first proximal operator in \eqref{eq:prox} depends on the regularising functional $R(\cdot)$, which is typically convex, so the computation can be done efficiently. For instance, it can be solved by quadratic programming \cite{nesterov1994interior} if $R(\cdot)$ is a Sobolev norm   and   with Chambolle's algorithm \cite{chambolle2004} if $R(\cdot)$ is the TV penalty.

The second proximal operator in \eqref{eq:prox} involves the high-dimensional constraint. However, the solution to \eqref{proxF} is simply the soft-thresholding operator applied to $\delta^{-1}w-KY$ with threshold $q_n\, \delta$. Altogether, the Chambolle--Pock algorithm applied to \eqref{MIND} is given in Algorithm~\ref{alg:ChambollePock}.

\begin{algorithm}
	\caption{Chambolle--Pock algorithm}
	\label{alg:ChambollePock}
	\begin{algorithmic}[1]
		\REQUIRE{$\delta,\tau>0$, $\theta\in[0,1]$, $k=0$, $(v_0,w_0)\in X\times Y$, stopping criterion}
		\vspace{0.25cm}
		\WHILE{stopping criterion not satisfied}
		\vspace{0.25cm}
\STATE	$w_{k+1}\leftarrow \big(I+\delta\, \partial F^{\star}\big)^{-1}(w_k+\delta\, K\, \widetilde{v}_k) $
		\vspace{0.25cm}

\STATE	$v_{k+1}\leftarrow \big(I+\tau\, \partial G\big)^{-1}(v_k-\tau\, K^*\, w_{k+1})$
		\vspace{0.25cm}

\STATE		$\widetilde{v}_{k+1}\leftarrow v_{N}+\theta\big(v_{k+1}-v_{k}\big)$
		\vspace{0.25cm}

\STATE
$k\leftarrow k+1$
		\vspace{0.25cm}

\ENDWHILE
	 \vspace{0.25cm}
		\STATE Return $(v_{k},w_k)$
	\end{algorithmic}
\end{algorithm}

  To  run the Chambolle--Pock algorithm, we need to choose the step sizes $\delta$ and $\tau$,   and   convergence of the algorithm is ensured whenever $\tau \delta \le \|K\|_{op}^{-2}$. Due to the different difficulty of the two subproblems, it is reasonable to choose $\delta\neq \tau$ and, in particular, to choose $\delta>\tau$. More precisely, the subproblem with respect to $F$ is more involved because of the large number of constraints, so this requires a much smaller step size, which by the Moreau's identity is equivalent to choosing a much larger $\delta$. We observe in practice that the choices $\delta=\|K\|_{op}^{-1}\sqrt{n}$ and $\tau=\|K\|_{op}^{-1}/\sqrt{n}$ yield good results  (see Section \ref{sec:num}). 

We remark that, in the Chambolle--Pock method applied to this problem, the high-dimensionality appears only in the very simple problem of soft-thresholding. This is a very convenient way of dealing with it and, as shown below,   makes the Chambolle--Pock-algorithm superior over the ADMM method.

\subsection{ADMM Method}

The alternating direction methods of multipliers (ADMM) can be seen as a variant of the augmented Lagrangian method  (see \cite{Pow69,Hes69} for classical references). To derive it, we introduce a slack variable $h\in \mathbb{R}^{\#\Lambda_n}$ and rewrite \eqref{MIND} it into the equivalent problem
\begin{equation*}
\underset{v\in\mathbb{R}^n, w\in\mathbb{R}^{\#\Lambda_n}}{\textup{argmin}} F(w)+G(v)\ \textup{ subject to } Kv=w.
\end{equation*}

By the convex duality theory, it is equivalent to find the saddle point of the augmented Lagrangian $L_{\lambda}(v,w;h)$, that is,
\begin{equation*}
\underset{v,w}{\textup{argmin}} \max_{h\in\mathbb{R}^{\#\Lambda_n}} F(w)+G(v)+\langle h, Kv-w\rangle+\frac{\lambda}{2}\|Kv-w\|^2
\end{equation*}
where $h\in\mathbb{R}^{\#\Lambda_n}$ is the Lagrangian multiplier    and   $\lambda>0$. As its name suggests, the ADMM
algorithm solves this saddle point problem alternately over $v,w$ and $h$ in a Gauß--Seidel fashion (i.e., successive displacement). The details are given in Algorithm \ref{alg:ADDM} below.

The usage of the ADMM for the problem \eqref{MIND} was first proposed by \citet{fmm12b}. One central difference compared to the Chambolle--Pock algorithm is that the ADMM splitting avoids the usage of $F^\star$. Instead, it deals with the high-dimensional constraint in an explicit way, which ultimately results in a slower performance.

From the steps performed by the ADMM in Algorithm \ref{alg:ADDM}, the first one (Line    2) involves the proximal operator of $R$ and can typically be dealt with with a standard algorithm (see the discussion in the Chambolle--Pock algorithm, Section \ref{sec:CP}). The second step (Line    3) is more challenging, as it involves solving the optimisation problem
\begin{equation}
w_{k}=\underset{w}{\textup{argmin}}\ \frac{\lambda}{2}\|w-(Kv_{k}+\lambda^{-1}h_{k-1})\|^2\ \textup{ subject to } \max_{\lambda\in\Lambda_n}|w_{\lambda}-\langle\phi_{\lambda},Y\rangle|\leq q_n.\label{Proj1}
\end{equation}

In other words, we have to find the orthogonal projection of the point $Kv_{N}+\lambda^{-1}h_{N-1}$ to the feasible set
\begin{equation*}
\big\{w\in\mathbb{R}^{\#\Lambda_n}\, |\, \max_{\lambda\in\Lambda_n}|w_{\lambda}-\langle\phi_{\lambda},Y\rangle|\leq q_n\big\}.
\end{equation*}

This set is the intersection of $2\#\Lambda_n$ half-spaces,   and   it is known to be non-empty (as it always contains $\{\langle Y, \phi_{\lambda}\rangle\, |\, \lambda\in\Lambda_n\}$). The projection problem \eqref{Proj1} can be solved by Dykstra’s algorithm \cite{dykstra1983algorithm, boyle1986method}, which converges linearly \cite{deutsch1994rate} (see \cite{birgin2005robust} for an efficient stopping rule).

Finally, it follows from Corollary 3.1 in \cite{deng2016global} that the ADMM has a linear convergence guarantee for the problem \eqref{primal}. 

\begin{algorithm}
	\caption{Alternating direction method of multipliers (ADMM)}
	\label{alg:ADDM}
	\begin{algorithmic}[1]
		\REQUIRE{data $Y\in\mathbb{R}^n$, step size $\lambda>0$, tolerance $\epsilon>0$, initial values $v_0,w_0,h_0$}
		\vspace{0.25cm}
		\WHILE{$\max\{ \|Kv_{k}-w_{k}\|,\|K(v_{k}-v_{k-1})\}>\epsilon$}
		\vspace{0.25cm}
\STATE\label{ADMMs1}	$v_{k}=\textup{argmin}_v\dfrac{\lambda}{2}\|Kv-(w_{k-1}-\lambda^{-1}h_{k-1})\|^2+G(v)$
		\vspace{0.35cm}
\STATE\label{ADMMs2} $w_{k}=\textup{argmin}_w\dfrac{\lambda}{2}\|w-(K v_{k}+\lambda^{-1} h_{k-1})\|^2 + F(w)$
		\vspace{0.35cm}
\STATE
$h_{k} = h_{k-1}+\lambda(Kv_{k}-w_k)$
		\vspace{0.25cm}

\STATE	$k\leftarrow k+1$
		\vspace{0.25cm}
		\ENDWHILE
		\vspace{0.25cm}
		\STATE Return $(v_{k},w_k)$
	\end{algorithmic}
\end{algorithm}

\subsection{Semismooth Newton Method}\label{ss:ssn}

Besides the optimality conditions \eqref{eq:opt}, the problem \eqref{MIND} can also be solved by the so-called Karush--Kuhn--Tucker conditions, which are necessary and sufficient as well. Using the notation of this section, the original problem \eqref{MIND} is equivalent to
\begin{equation}
\min_{v \in \mathbb R^n} R \left(v\right) \qquad\text{such   that}\qquad q_n - Kv + KY \geq 0, q_n - KY + Kv \geq 0.\label{SSNform}
\end{equation}

This is a convex optimisation problem with linear inequality constraints,   and   if we introduce $B: \mathbb R^n \to \mathbb R^{2\# \Lambda_n}$ as $v \mapsto B v$ with $B = (K^\top, -K^\top)^\top$, then a vector $v$ is a solution to \eqref{SSNform} if and only if there exists a vector of Lagrange multipliers $\lambda \in \mathbb R^{2\#\Lambda_n}_{\geq 0}$ such that
\begin{subequations}\label{eq:KKT}
\begin{align}
\partial R \left(v\right) &\ni -B^\top\lambda, \label{eq:KKT1}\\
q_n - Bv + BY &\geq 0,\\
\lambda_i \left(q_n - Bv + BY\right)_i & = 0 \quad\text{for all}\quad 1 \leq i \leq 2 \#\Lambda_n.
\end{align}
\end{subequations}

The latter two conditions can be reformulated as
\begin{equation}\label{eq:lambda}
\lambda_i = \max\left\{0, \lambda_i +c\left(Bv - BY - q_n\right)_i \right\} \quad\text{for all}\quad 1 \leq i \leq 2 \#\Lambda_n.
\end{equation}

The immediately visible advantage of this formulation over the original problem is that the inequality constraints have been transformed into equations. Now, suppose for a moment that the functional $R$ is twice differentiable. Then,  $\partial R(v)$ is single valued and \eqref{eq:KKT1} is a differentiable equation for $v$ and $\lambda$. It seems a natural approach to solve the corresponding system of Equations \eqref{eq:KKT1} and \eqref{eq:lambda} via an analog of Newton's method. On the other hand, even if $R$ was twice differentiable, the $\max$ function in \eqref{eq:lambda} is not differentiable in the classical sense. The application of Newton's method is however still possible, as the $\max$ function turns out to be semismooth (see Definition 2.5 in \cite{hintermueller}). For a general operator $B$, it is however not clear if the overall system \eqref{eq:KKT1} and \eqref{eq:lambda} can also be described as finding the root of a semismooth function. Therefore, one introduces a regularisation parameter $\beta \in \left(0,1\right)$ and replaces \eqref{eq:lambda} by the regularised equation $\lambda_i = \beta \max\left\{0, \lambda_i +c\left(Bv - BY - q_n\right)_i \right\}$ for all $1 \leq i \leq 2 \# \Lambda_n$, which is in turn equivalent to
\begin{equation}\label{eq:lambda2}
\lambda_i = \max\left\{0, \frac{\beta c}{1-\beta} \left(Bv - BY -q_n\right)_i \right\}\quad\text{for all}\quad 1 \leq i \leq 2 \#\Lambda_n.
\end{equation}

This system is now explicit in $\lambda$ and yields in combination with \eqref{eq:KKT1} the overall system
\begin{equation}\label{eq:ssn}
\partial R \left(v\right) \ni -\frac{1}{\delta} B^\top\max\left\{0, \left(Bv - BY -q_n\right) \right\}
\end{equation}
with the new regularisation parameter $\delta := {(1-\beta)}/{(\beta c)}\in \left(0,\infty\right)$. This system can---for twice differentiable $R$ and under appropriate assumptions on $B$ satisfied in our example---now be shown to be semismooth (cf. \cite{hk06}). Furthermore, for $\delta \searrow 0$, the solution of the regularised system \eqref{eq:ssn} converges towards a solution of the original system \eqref{eq:KKT}. This follows from the fact that \eqref{eq:lambda2} is in fact the Moreau--Yosida regularisation of the second optimality condition \eqref{eq:opt2}. In practice, this limiting process is realised by a path-continuation scheme, sustaining the superlinear convergence behaviour.

For differentiable $R$, such as the Sobolev seminorm $R(v)=|v|_{H^s}$, the implementation of the semismooth Newton method with path-continuation is now straightforward:  the overall system to be solved can now be written as
\begin{equation}\label{eq:newton}
T_{\delta}(v)=0, \ \textup{ where } T_{\delta}=\partial |v|_{H^s}+\delta^{-1}B^T\max\{0, Bv - BY-q_n\}.
\end{equation}

Denote by $\mathcal{D}_k[T_{\delta}]$ the generalised derivative of the functional at the position $u_k$. Then,  we initialise the iteration at $\delta_0 > 0$ and $u_{0,\delta}$ and solve the linear equations
\begin{equation*}
\mathcal{D}_k[T_{\delta}]u_{k+1,\delta}=\mathcal{D}_k[T_{\delta}]u_{k,\delta}+T_{\delta}(u_{k,\delta}) \ \ \ \textup{ for }k\geq 0
\end{equation*}
iteratively until a stopping criterion is satisfied. Then,  the parameter $\delta$ is decreased by a fixed factor (say $1/2$) and the iteration is started again with $u_{k,\delta}$ as the initial guess. This continuation process is stopped until a global error criterion is reached, which is formulated in terms of the number and magnitude of the violated constraints  (see Algorithm \ref{alg:SSN}). {Note that, for $\delta > 0$, the underlying minimisation problem is strictly convex (this is an immediate property of the Moreau--Yosida regularisation, cf. \cite{clason2018total}) and the subdifferential is Lipschitz continuous with Lipschitz constant $\sim$$1/\delta$,   and   hence the Newton iteration for \eqref{eq:newton} will converge for all initial guesses in a ball with radius $\sim$$\delta$ around $0$. Hence, whenever the inner iteration in Algorithm \ref{alg:SSN} does not converge, it should be re-started with a larger value of $\delta$. However, a larger value of $\delta$ clearly prolongs the run-time of Algorithm \ref{alg:SSN},   and   hence the initial value $\delta$ should not be too large.}

In the case of a non-differentiable $R$ such as the TV-seminorm, we introduce another regularisation for $R$, resulting in its Huber regularisation
\[
TV_\beta \left(v\right) = \int \min\{\beta^{-1}\,\left|\nabla v\right|^2,\left|\nabla v\right|\}\, d x.
\]

This functional is differentiable for any $\beta > 0$,   and   to compute the limit $\beta \searrow 0$ we again apply a path-continuation strategy (cf. \cite{clason2018total}). In this case, the superlinear convergence behaviour is sustained. However, we remark that the path-following routine for two parameters ($\delta$ and $\beta$) turns out to be unstable in practice. Instead, it might be desirable to look for the Moreau--Yosida regularisation of \textit{both} the constraint and $R$, that is, to regularise the optimality conditions \eqref{eq:KKT} in one step with just one parameter. We leave this idea to be pursued in future work.

Finally, we remark that this superlinear convergence of the semismooth Newton method is a huge theoretical advantage over the Chambolle--Pock and the ADMM algorithms. In practice, however, the situation is more complex, as the convergence speed of Newton's method depends strongly on the initialisation. {In Table \ref{tab:Comparison}, we provide a summary of the comparison between the three methods.}

\begin{algorithm}
	\caption{Semismooth Newton method for $R(v)=|v|_{H^s}$}
	\label{alg:SSN}
	\begin{algorithmic}[1]
		\REQUIRE{data $Y\in\mathbb{R}^n$, step size $\Delta\delta>0$, tolerance $\epsilon, \delta_{\min},\rho_{min},r_{min}>0$, initial value $\delta$}
		\vspace{0.25cm}
		\STATE $v_{old}\leftarrow 0\in\mathbb{R}^n$ 
		\vspace{0.25cm}
		\STATE $ratio,\ res\leftarrow 1$ 
		\vspace{0.25cm}
		\WHILE{$\delta>\delta_{min}$}
		\vspace{0.25cm}
		\WHILE{$ratio>\rho_{min}$ or $res>r_{min}$}
		\vspace{0.25cm}
		\STATE $v_{new}\leftarrow$ solution to equation: $\ \nabla T_{\delta}(v_{old})v_{new}=\nabla T_{\delta}(v_{old})v_{old}-T_{\delta}(v_{old})\ $ with tolerance $\epsilon$
		\vspace{0.35cm}
		\STATE $ratio\leftarrow \frac{\#\{q_n-Bv_{new}+BY<0\}}{2\#\Lambda_n}$ 
		\vspace{0.35cm}
		\STATE $res\leftarrow \frac{\sqrt{ \|\max\{Kv_{new}-KY-q_n,0\}\|^2+\|\max\{KY-q_n-Kv_{new},0\}\|^2}}{\|v_{new}\|}$ 
		\vspace{0.25cm}
		\ENDWHILE
		\vspace{0.25cm}
		\STATE $v_{old}\leftarrow v_{new}$
		\vspace{0.25cm}
		\STATE $\delta\leftarrow\delta\cdot \Delta\delta$
		\vspace{0.25cm}
		\ENDWHILE
		\vspace{0.25cm}
		\STATE Return $v_{old}$
	\end{algorithmic}
\end{algorithm}

\begin{table}
\caption{{Summary of the comparison of the Chambolle--Pock, ADMM and semismooth Newton algorithms.}}
\label{tab:Comparison}
\centering
{\footnotesize
\begin{tabular}{cccc}
\toprule
 & \textbf{\specialcell{Dependence on \\ Initialisation}}	& \textbf{\specialcell{Theor. Convergence \\ Speed}}	& \textbf{\specialcell{Practical \\ Performance}}\\
\midrule
 Chambolle--Pock & no & linear & \specialcell{Good for smooth \\ and nonsmooth $R$} \\\midrule 
 ADMM & no & linear & Too slow \\ \midrule
 Semismooth Newton & yes & superlinear & \specialcell{Good for smooth $R$ \\ Unstable otherwise} \\ 
\bottomrule
\end{tabular}}
\end{table}

\section{Numerical study}\label{sec:num}

In this section, we compare the practical performance of the Chambolle--Pock,   ADMM and   semismooth Newton algorithms in computing MIND    and   demonstrate the empirical performance of MIND with respect to various choices of dictionaries. We refer to, e.g., \cite{CandesGuo,fmm12a,fmm12b,fmm13,grasmair2015,khkl16,del2018frame} for further numerical examinations of MIND. We select in particular Sobolev $H^1$ and TV seminorms as the regularisation functional, with the former differentiable but the latter non-differentiable   and   the $50\%$-quantile of $\|\sigma\epsilon\|_{\text{MS}}$ in \eqref{Residual} as $q_n$ for MIND. Concerning measure of image quality, we consider the peak signal-to-noise ratio (PSNR \cite{hore2010image}), the structural similarity index measure (SSIM  \cite{wang2004image}) and the visual information fidelity (VIF  \cite{sheikh2006image}) criteria. The implementation in MATLAB for the Chambolle--Pock algorithm, together with code that reproduces all the following numerical examples, is available at \url{https://github.com/housenli/MIND}.

\subsection{Comparison of Three Algorithms}

\begin{figure}
 \centering
 \begin{tabular}{cc}
 \includegraphics[width=0.3\textwidth]{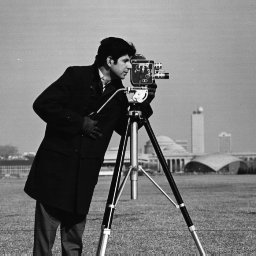} & \includegraphics[width=0.3\textwidth]{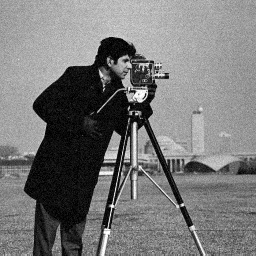} \\
 \subcap{(\textbf{a}) Truth} & \subcap{(\textbf{b}) Noisy, PSNR = 29.5} 
 \end{tabular}
 \caption{Image ``cameraman'' and its noisy version with SNR = 30.}
 \label{f:cm}
\end{figure}

\begin{figure}
 \centering
 \begin{tabular}{ccc}
 \includegraphics[width=0.3\textwidth]{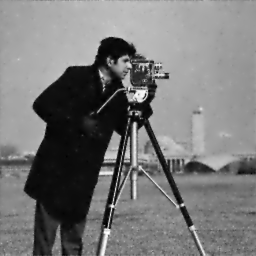} & \includegraphics[width=0.3\textwidth]{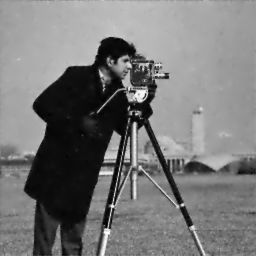} & \includegraphics[width=0.3\textwidth]{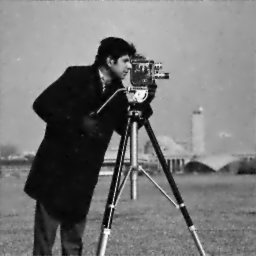}\\
 \subcap{(\textbf{a}) Chambolle--Pock} & \subcap{(\textbf{b}) ADMM} & \subcap{(\textbf{c}) Semismooth Newton}
 \end{tabular}
 \caption{Limit solutions of all three algorithms for MIND with $H^1$ (PSNR = 30 for all) regularisation after long iterations.}
 \label{f:cmgoa}
\end{figure}

\begin{figure}
 \centering
 \begin{tabular}{cc}
 \includegraphics[width=0.3\textwidth]{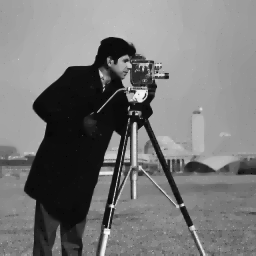} & \includegraphics[width=0.3\textwidth]{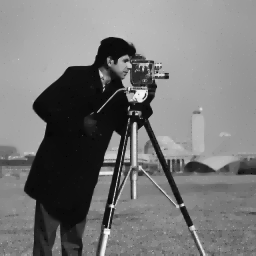} \\
 \subcap{(\textbf{a}) Chambolle--Pock} & \subcap{(\textbf{b}) ADMM} 
 \end{tabular}
 \caption{Limit solutions of Chambolle--Pock and ADMM algorithms for MIND with TV (PSNR = 30.6 for both) regularisation after long iterations. As described in the main text, the semismooth Newton method showed an unstable behaviour in combination with TV regularisation,   and   hence the result is not documented here.}
 \label{f:cmgob}
\end{figure}

Here, we compare the convergence speed of the Chambolle--Pock,   ADMM and   semismooth Newton algorithms in solving \eqref{MIND} or equivalently \eqref{primal}. We stress that the relative performance of the three algorithms remains similar over different settings. Thus, as a particular example, we choose the ``cameraman'' image ($256 \times 256$ pixels) as the function $f$ in model \eqref{model}   and   the indicator functions of dyadic (partition) system of cubes (cf. Definition 2.2 in \cite{grasmair2015}), which consists of cubes
\begin{equation}\label{eq:dyadic}
 \left[i 2^{-\ell}, \,(i+1)2^{-\ell}\right]\times\left[j 2^{-\ell}, \,(j+1)2^{-\ell}\right] \quad \subseteq \quad [0,1]^2 \qquad \text{ for all possible } i, j, \ell \in \mathbb{N}
\end{equation}
as the dictionary $\{\phi_{\lambda}| \lambda\in\Lambda_n\}$. The noise level $\sigma$ is assumed to be known and is chosen such that the signal-to-noise ratio (SNR) $\max_{i = 1,\ldots,n}|f(x_i)|/\sigma = 30$, see Figure \ref{f:cm}. Besides the aforementioned image quality measures, we employ objective values $R(g_k)$, (relative) constraint gaps $\bigl(\max_{\lambda\in\Lambda_n} |\langle \phi_{\lambda},g_k-Y\rangle|-q_n\bigr)/q_n$,   and   the distance $\|{g_k -g_{\infty}}\|$ to the limit solution $g_{\infty}$ to examine the evolution of iterations $g_k$\,. The limit solution $g_{\infty}$ is obtained for each algorithm after a large number of iterations, that is  $3\times 10^4$ outer iterations for the Chambolle--Pock, $3 \times 10^3$ outer iterations for the ADMM,   and   the largest possible number of iterations until which the path-continuation scheme remains stable for the semismooth Newton method (in the considered setup, $29$ outer iterations in case of $H^1$ regularisation). As mentioned   in Section \ref{ss:ssn}, the semismooth Newton method is unstable for the case of TV regularisation, which is thus not reported here. All limit solutions $g_\infty$ are shown in Figures \ref{f:cmgoa} and \ref{f:cmgob}    and   are visually quite the same in each figure (while MIND with TV regularisation leads to a slightly better result than that with $H^1$ regularisation). This indicates that all three algorithms are able to find the global solution of \eqref{MIND} to a desirable accuracy. 

\unskip

\begin{figure}
\centering
 \begin{tabular}{cc}
 \includegraphics[width=0.43\textwidth]{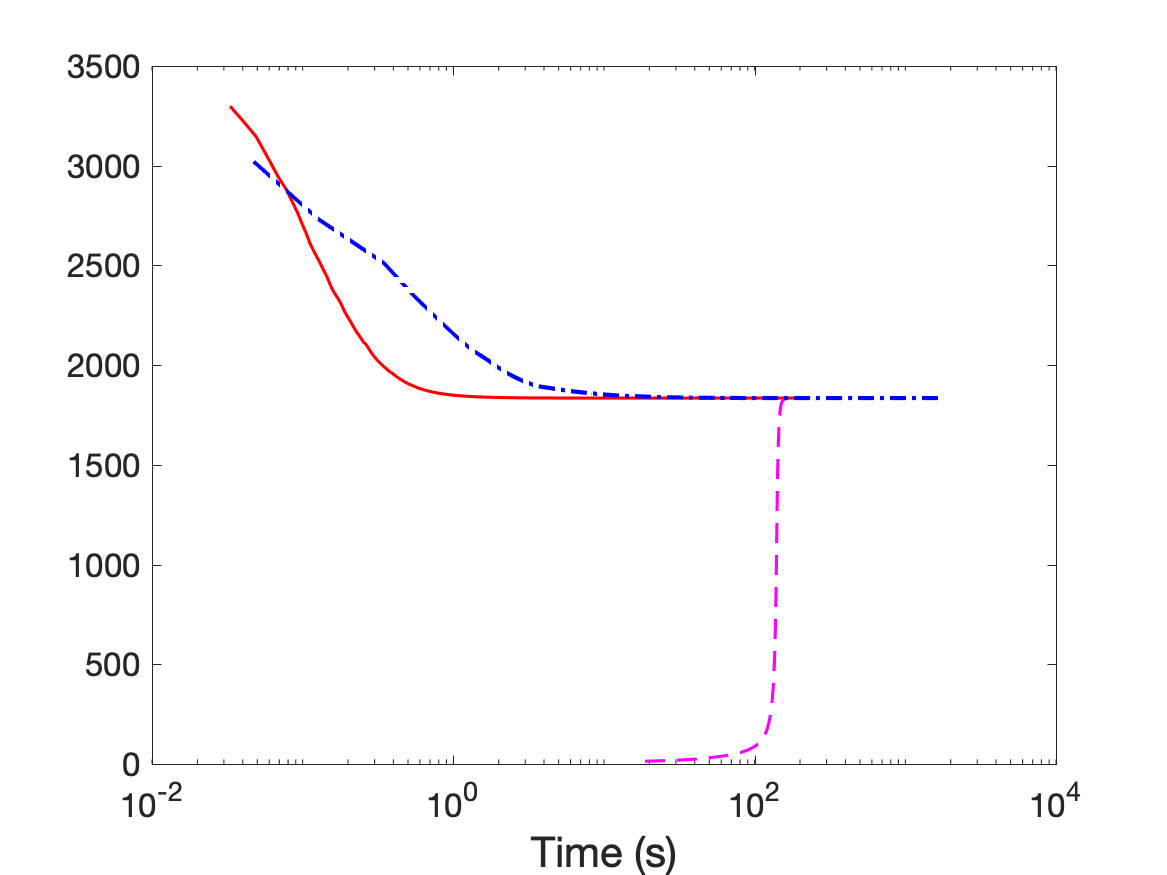} & \includegraphics[width=0.43\textwidth]{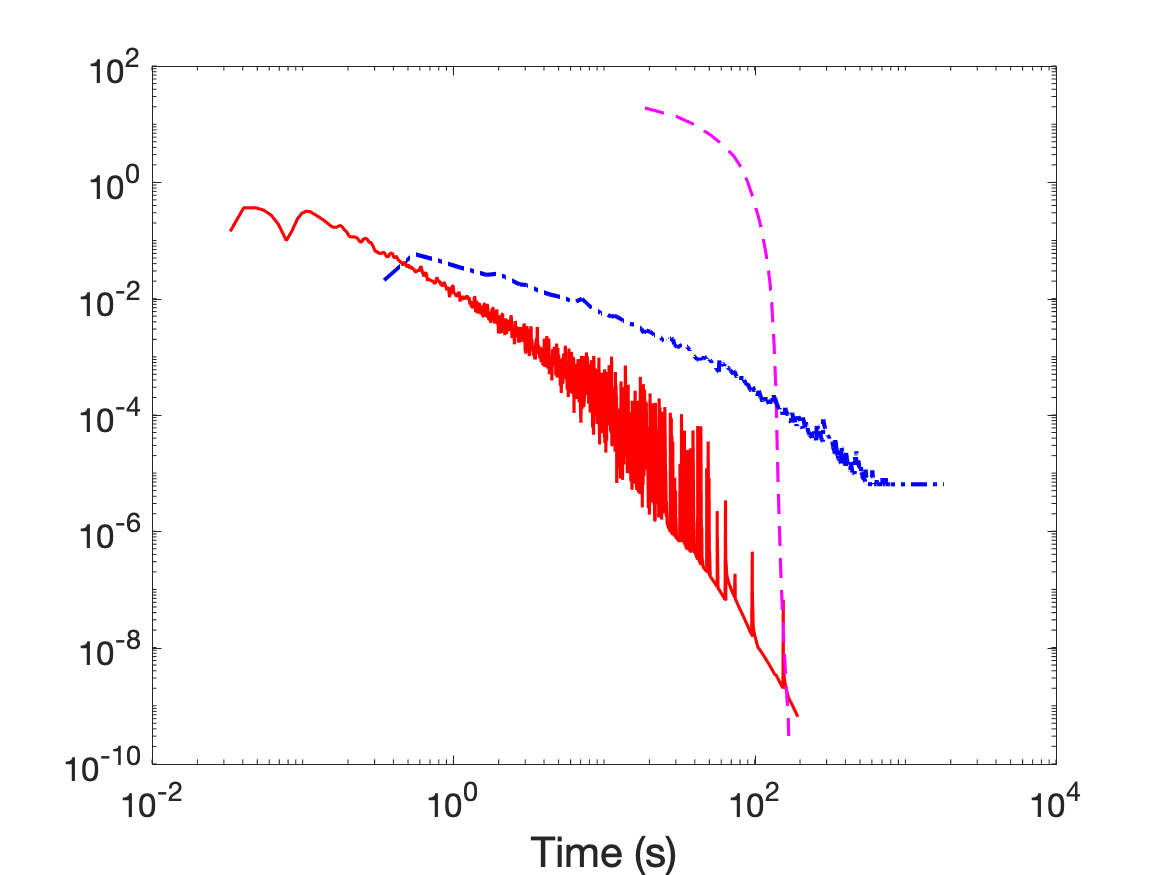} \\
 \subcap{(\textbf{a}) Objective value}
 & \subcap{(\textbf{b}) Constraint gap}\\
 \includegraphics[width=0.43\textwidth]{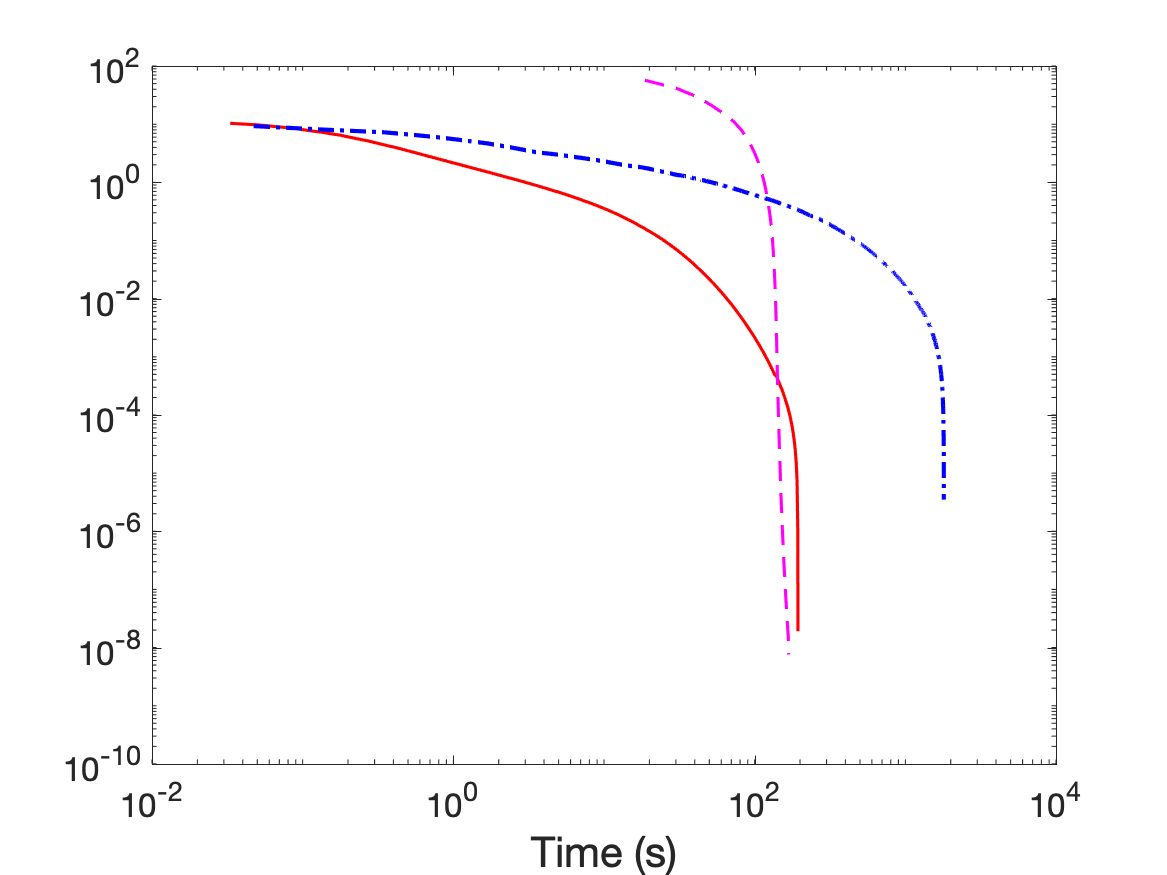}& \includegraphics[width=0.43\textwidth]{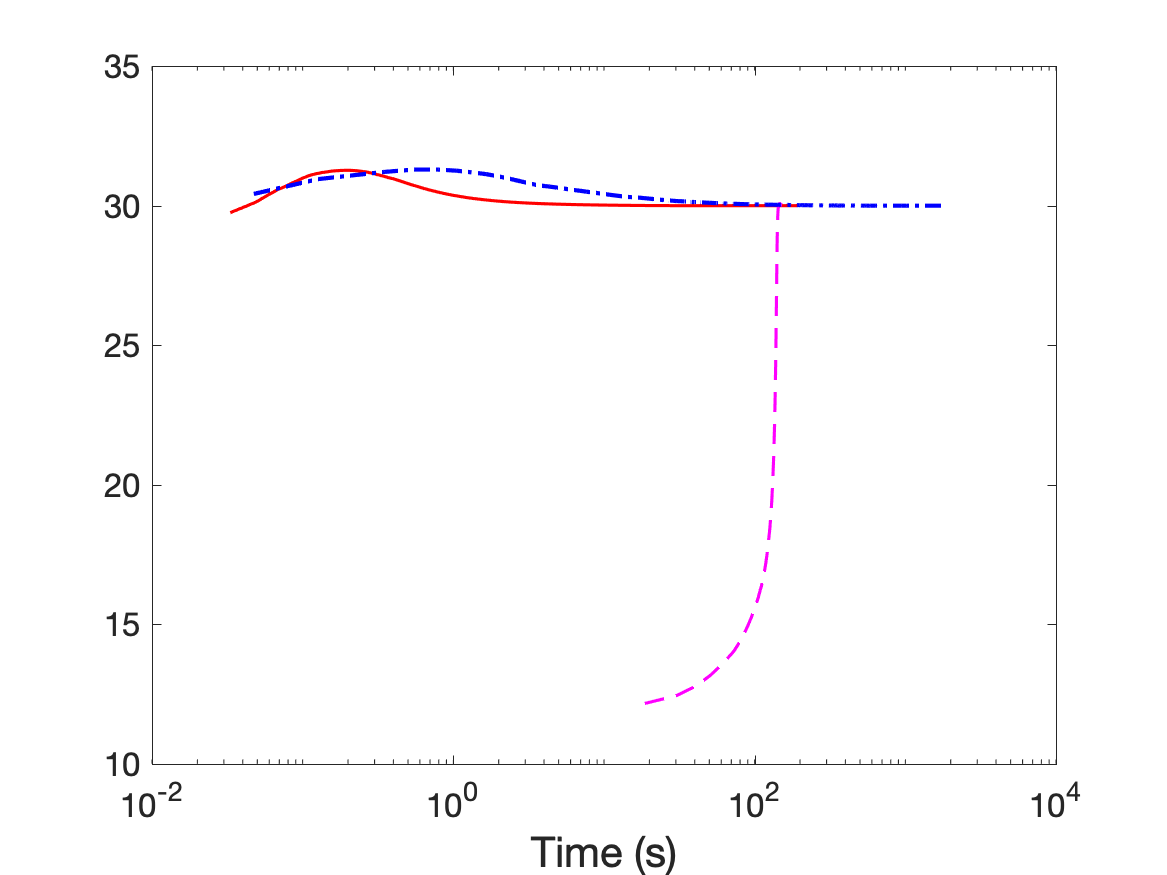}\\
 \subcap{(\textbf{c}) Distance to limit} & \subcap{(\textbf{d}) PSNR}\\
 \includegraphics[width=0.43\textwidth]{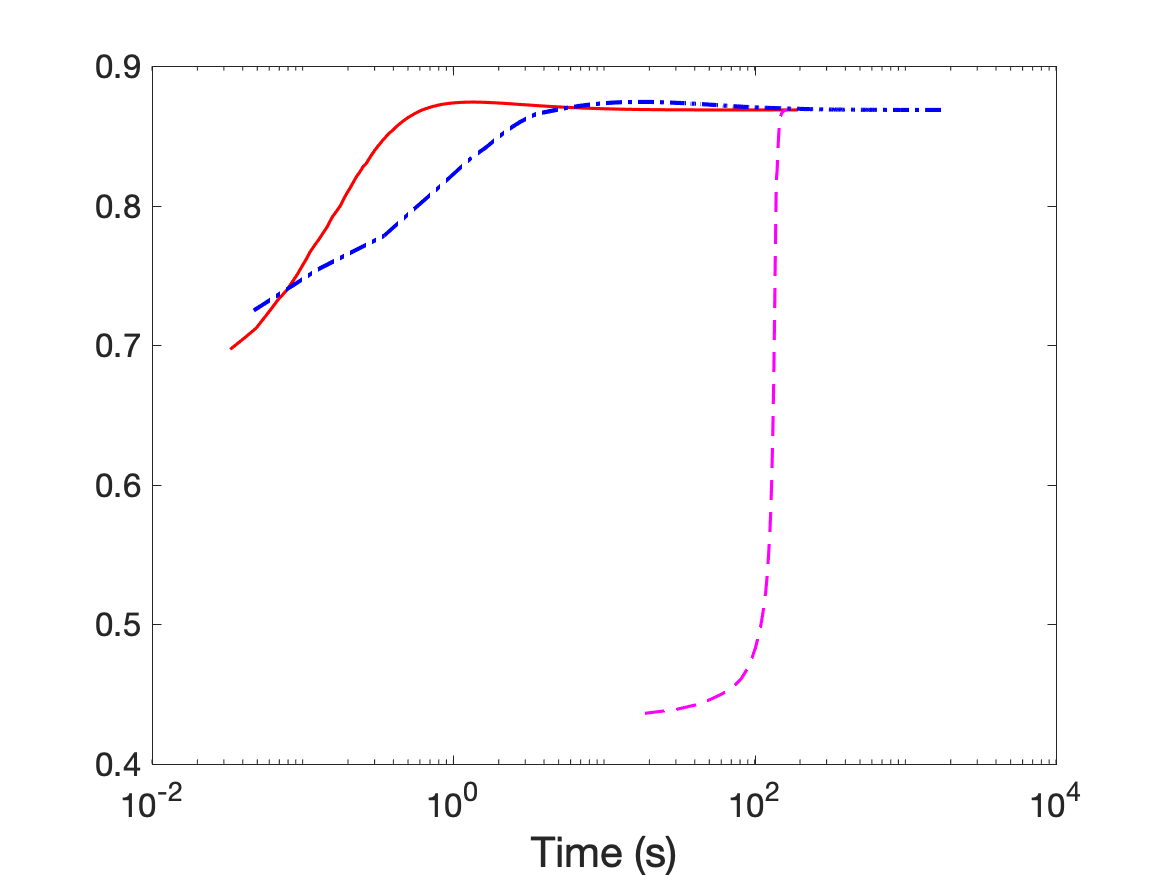} & \includegraphics[width=0.43\textwidth]{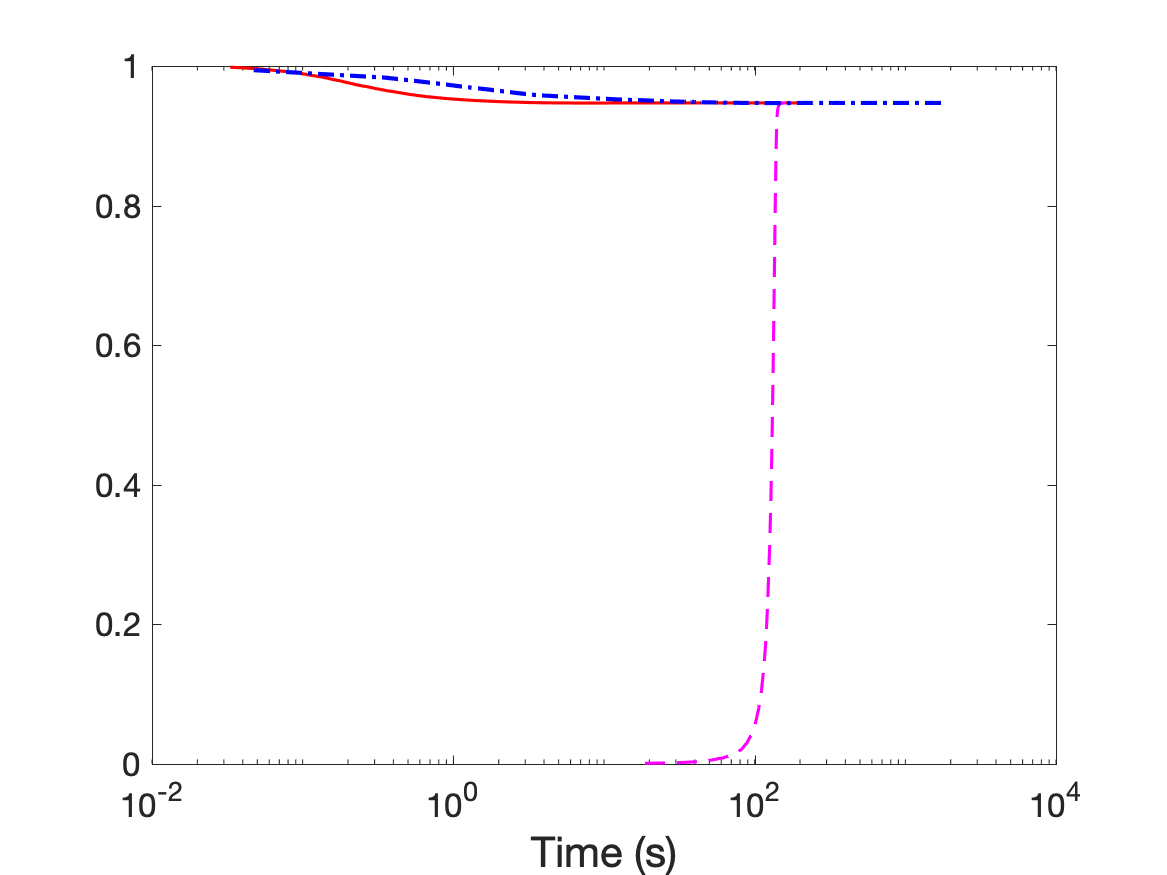}\\
 \subcap{(\textbf{e}) SSIM} & \subcap{(\textbf{f}) VIF}
 \end{tabular}
 \caption{Performance of Chambolle-{Pock} (\textcolor{red}{line}), ADMM (\textcolor{blue}{dash-dot}) and semismooth Newton (\textcolor{purple}{dash}) algorithms for MIND with $H^1$ regularisation over time.}
 \label{f:qmh1}
\end{figure}

\begin{figure}
 \centering
 \begin{tabular}{cc}
 \includegraphics[width=0.43\textwidth]{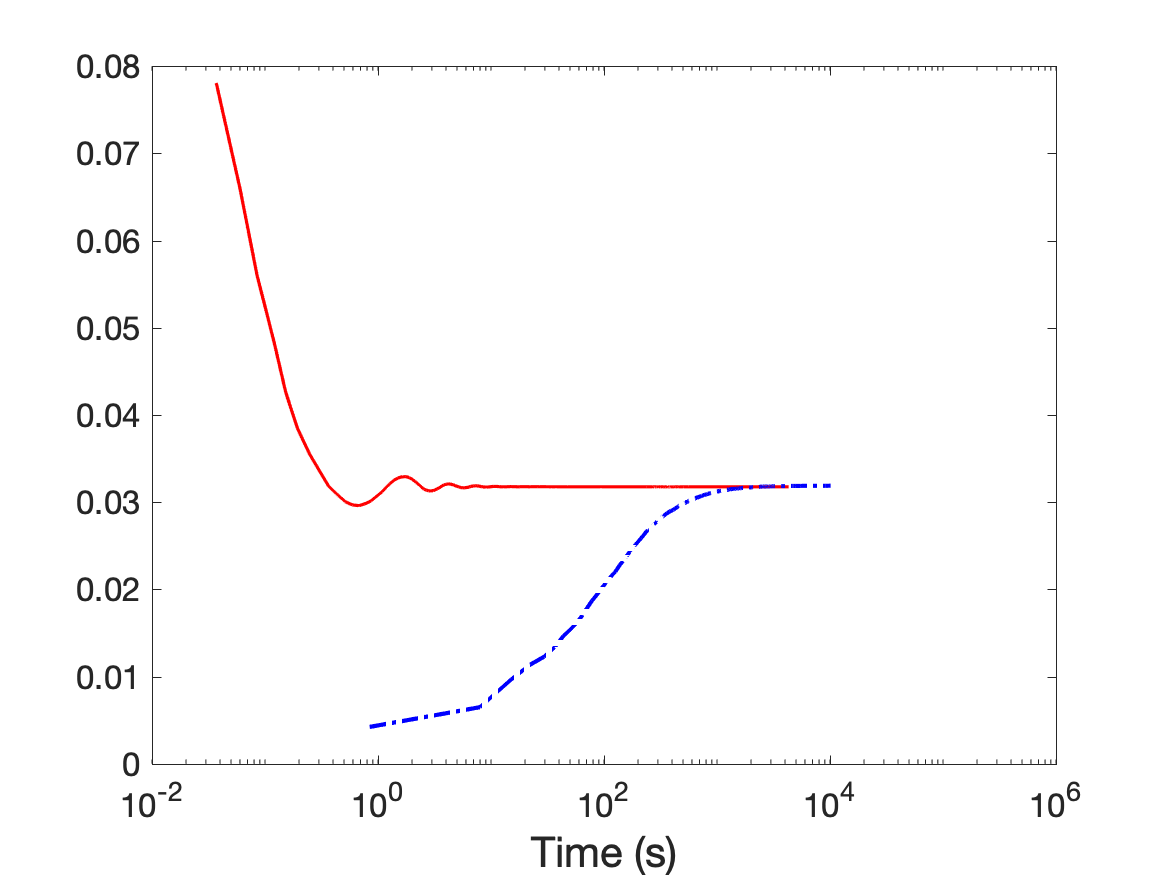} & \includegraphics[width=0.43\textwidth]{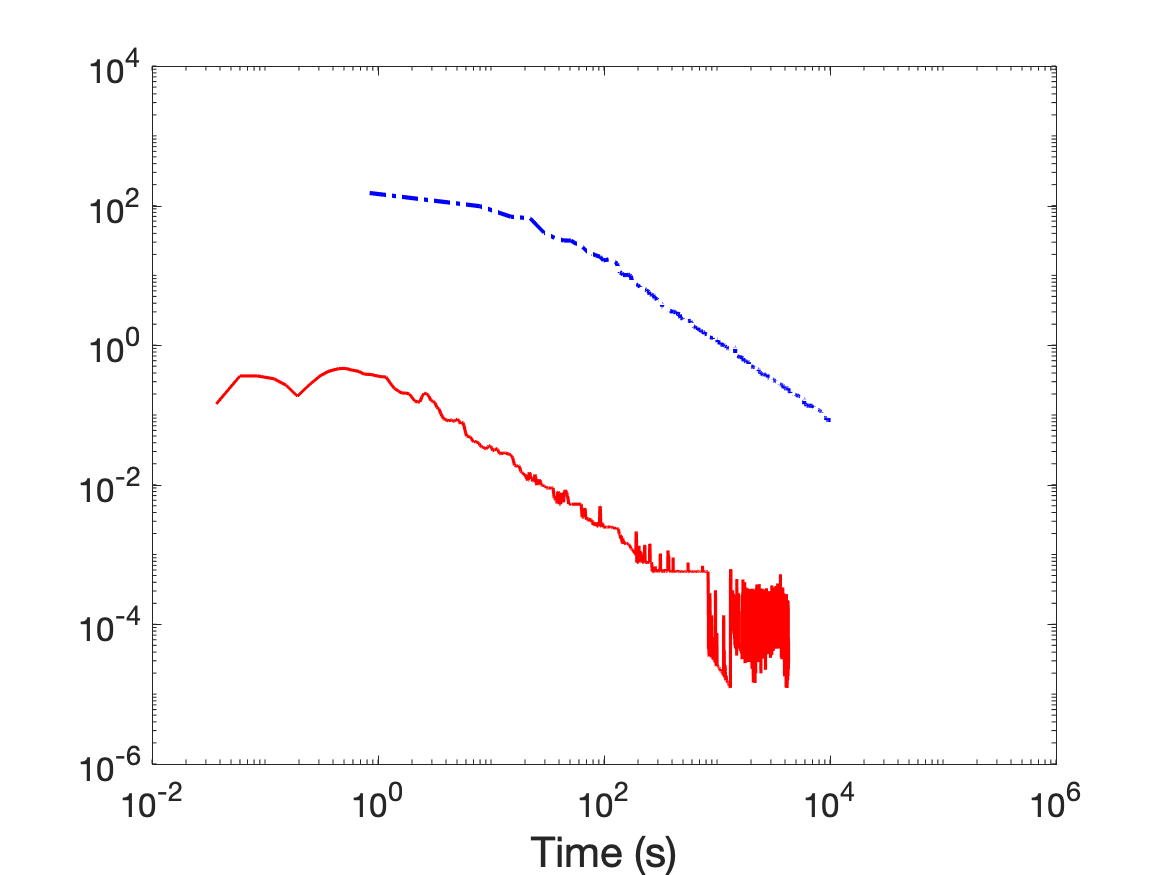} \\
\subcap{ (\textbf{a}) Objective value}
 & \subcap{(\textbf{b}) Constraint gap}\\
 \includegraphics[width=0.43\textwidth]{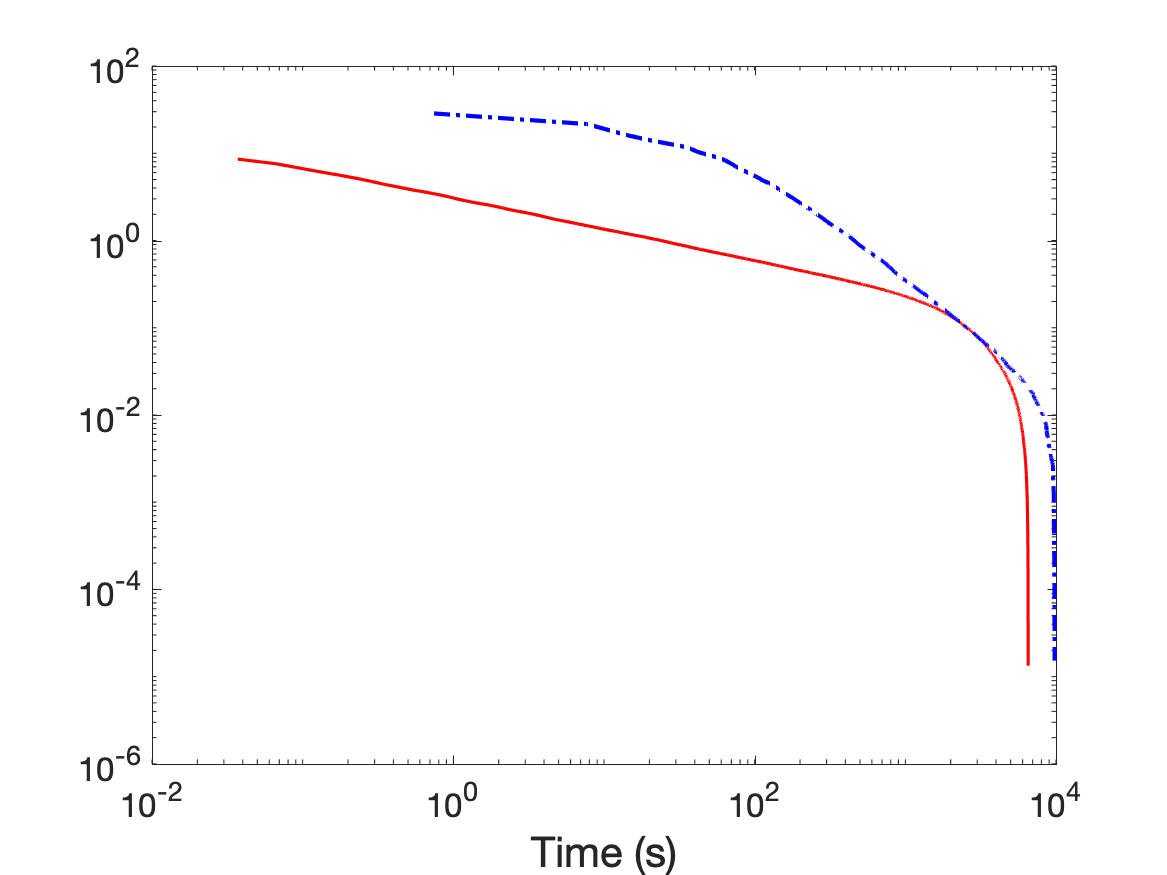}& \includegraphics[width=0.43\textwidth]{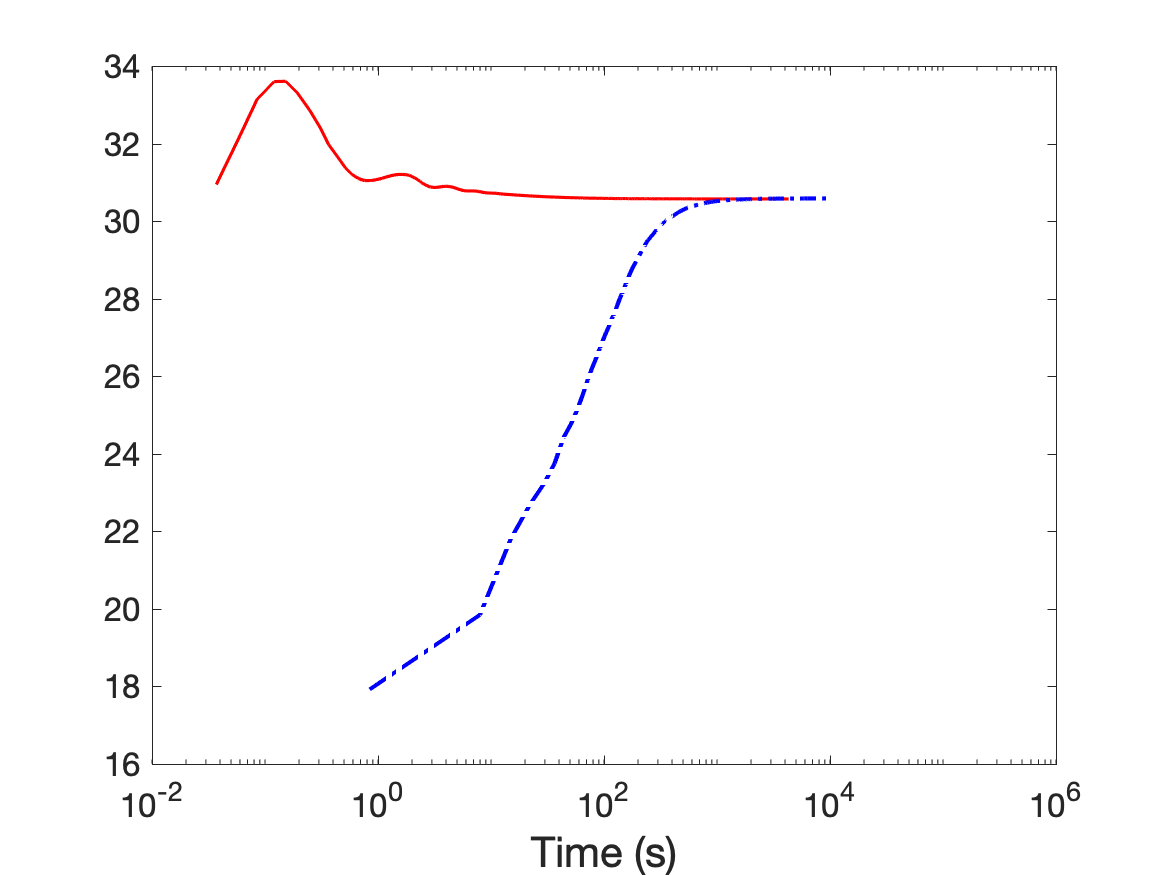}\\
 \subcap{(\textbf{c}) Distance to limit} & \subcap{(\textbf{d}) PSNR}\\
 \includegraphics[width=0.43\textwidth]{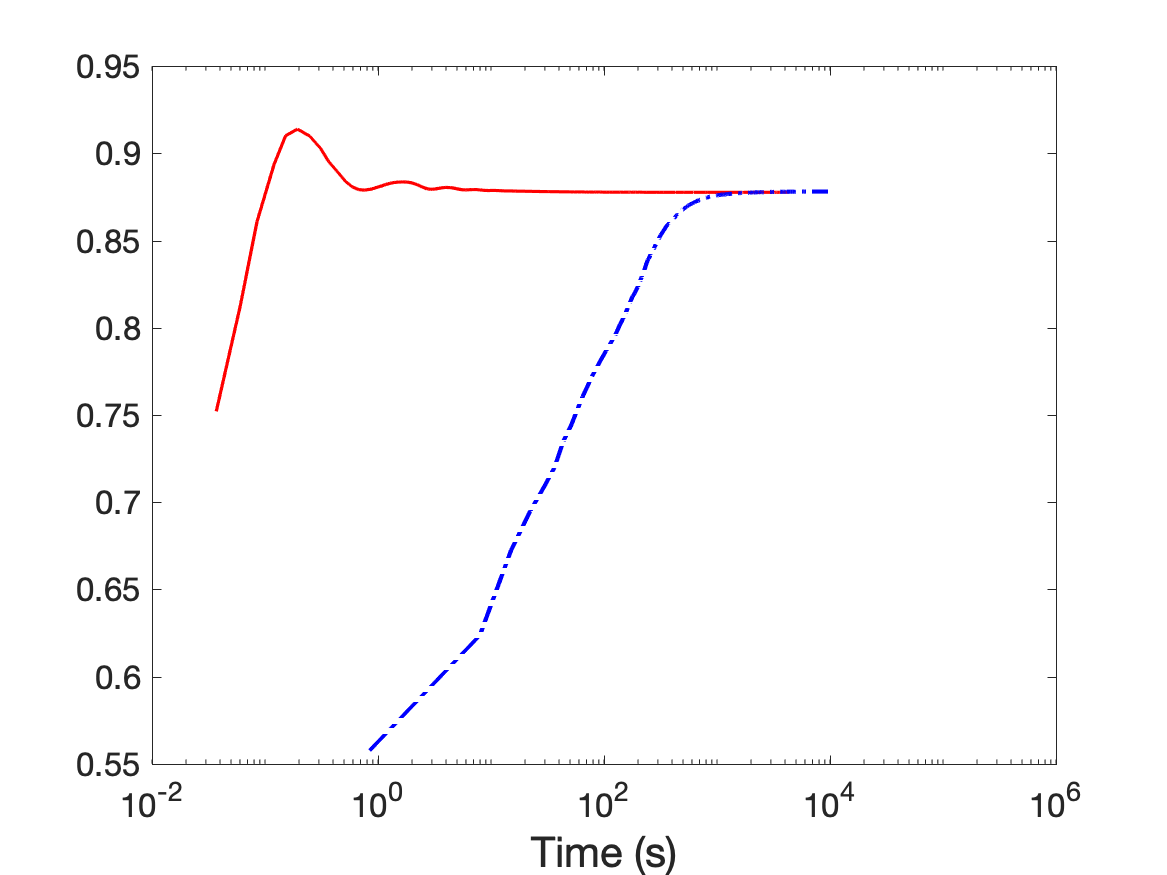} & \includegraphics[width=0.43\textwidth]{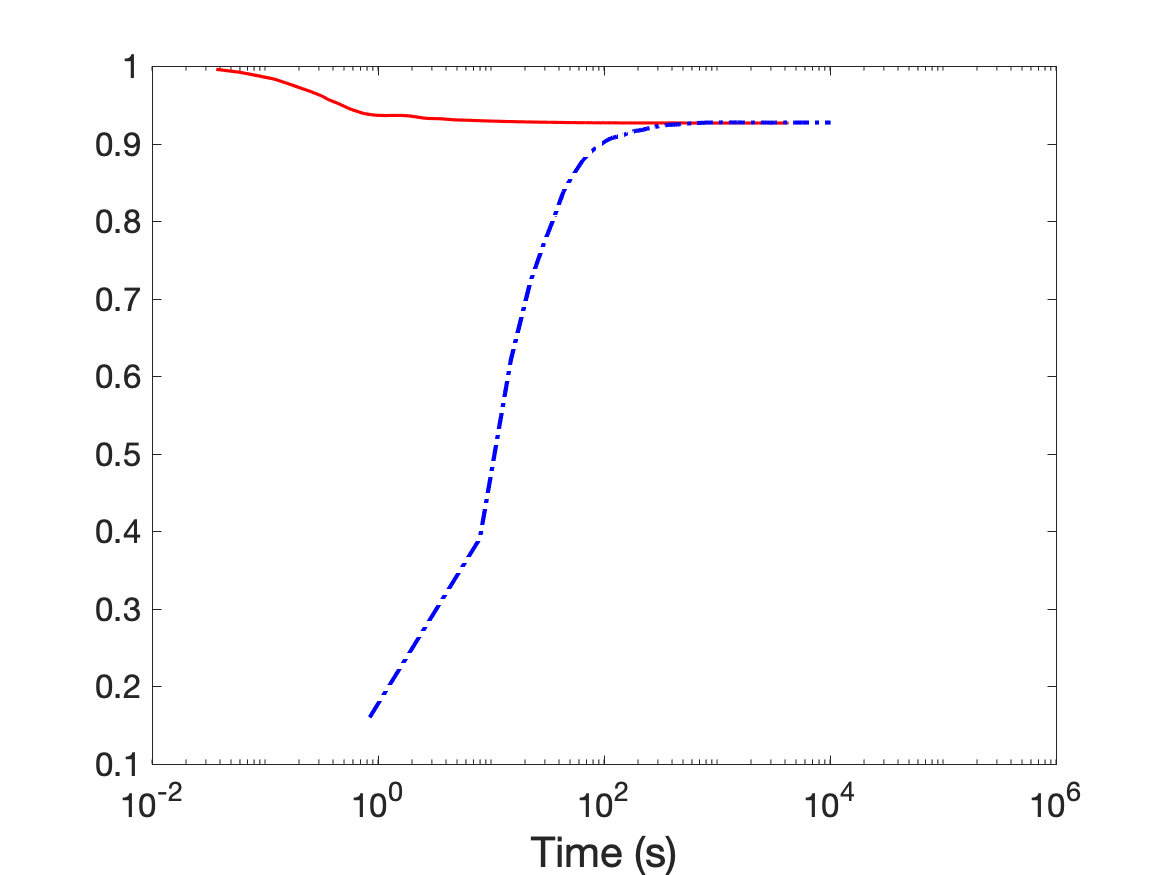}\\
 \subcap{(\textbf{e}) SSIM} & \subcap{(\textbf{f}) VIF}
 \end{tabular}
 \caption{Performance of Chambolle--Pock (\textcolor{red}{{line}}) and ADMM (\textcolor{blue}{dash-dot}) algorithms for MIND with TV regularisation over time. }
 \label{f:qmtv}
\end{figure}

To reduce the burden of practitioners on parameter tuning, we set the parameters in the Chambolle--Pock (Algorithm \ref{alg:ChambollePock}) by default as $\theta = 1$, $\delta=\|K\|_{op}^{-1}\sqrt{n}$ and $\tau=\|K\|_{op}^{-1}/\sqrt{n}$ in all settings. 
We note that the performance of the Chambolle--Pock, which is reported below, can be further improved by fine tuning $\theta$ and $\delta$    and   possibly also by preconditioning \cite{pock2011diagonal}, for every choice of regularisation functional and dictionary. In contrast, the parameters of the ADMM and the semismooth Newton algorithms are fine tuned towards the best performance for each case. More precisely, for   ADMM, we choose $\lambda = 50$ in the case of $H^1$ and $\lambda = 0.5$ in the case of TV,   and,   for the semismooth Newton, we choose $\delta = 1/3$ and $\Delta\delta =1/3$ in the case of $H^1$. 
The performance of three algorithms over time is shown in Figures \ref{f:qmh1} and \ref{f:qmtv} for $H^1$ and TV regularisation, respectively. In both cases, the Chambolle--Pock clearly outperformed the ADMM with respect to all considered criteria. The semismooth Newton algorithm, as is ensured by the theory, exhibited a faster convergence rate than the other two, but this happened only at a very late stage. Moreover, the path-continuation scheme is sensitive to the choice of parameters    and   makes it difficult to decide the right stopping stage in general. In summary, from a practical perspective, the overall performance of the Chambolle--Pock algorithm   was found to be superior compared to the other algorithms. We speculate, however, that a hybrid combination of the Chambolle--Pock and the semismooth Newton algorithm (switching to the semismooth Newton algorithm after a burn in period using the Chambolle Pock algorithm) might lead to further improvement. 

\subsection{Comparison of Different Dictionaries}\label{ss:dict}

We next investigate the choice of dictionary $\{\phi_{\lambda}| \lambda\in\Lambda_n\}$ and its impact on the practical performance. Five different dictionaries are considered. The first consists of the indicator functions of dyadic cubes defined in \eqref{eq:dyadic},   and   the second is composed of the indicator functions of small cubes of edge length $\le 30$, i.e.,
$$
\frac{1}{n}\,[i,i+\ell] \times [j,j+\ell] \quad \subseteq \quad [0,1]^2 \qquad \text{for } \ell =1, \ldots, 30, \text{ and all possible } i,j \in \mathbb{N}\,.
$$
The third is the system of tensor wavelets, in particular, the Daubechies' symlets with six  vanishing moments. The fourth is the frame of curvelets    and   the last is that of shearlets, both of which are constructed using default parameters in packages CurveLab (\url{http://www.curvelet.org}) and ShearLab3D (\url{http://shearlab.math.lmu.de}), respectively.
As shown in Figures \ref{f:cmgoa} and \ref{f:cmgob}, different algorithms lead to almost the same final result, we only report the Chambolle--Pock algorithm, due to its empirically fast convergence.

\begin{figure}
 \centering
 \begin{tabular}{ccc}
 \includegraphics[width=0.3\textwidth]{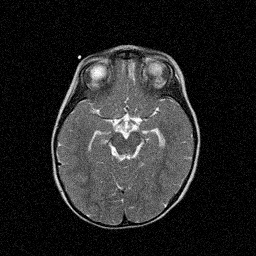} 
 &
 \includegraphics[width=0.3\textwidth]{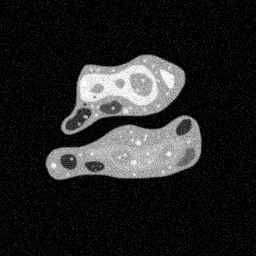} 
 & 
 \includegraphics[width=0.3\textwidth]{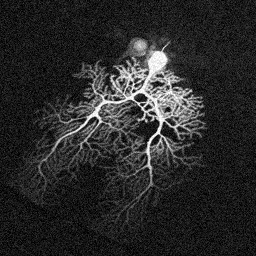} 
 \\
 \subcap{(\textbf{a}) Brain} & \subcap{(\textbf{b}) Cell} & \subcap{(\textbf{c}) BIRN}
 \end{tabular}
 \caption{Noisy images of ``brain'', ``cell'' and ``BIRN'' with SNR = 20 and PSNR = 26.}\label{f:noise}
\end{figure}

\begin{figure}
 \centering
 \begin{tabular}{ccc}
 \includegraphics[width=0.3\textwidth]{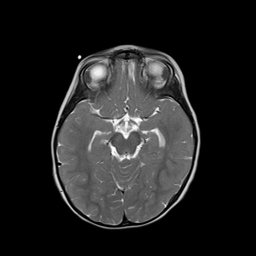} 
 &
 \includegraphics[width=0.3\textwidth]{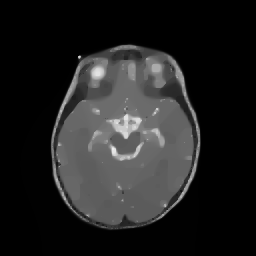} 
 & 
 \includegraphics[width=0.3\textwidth]{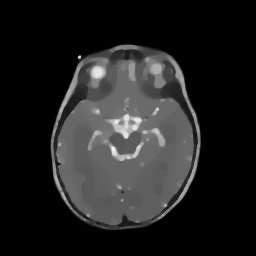} 
 \\
 \subcap{(\textbf{a}) Truth} & \subcap{(\textbf{b}) Dyadic cubes, PSNR = 27.5} & \subcap{(\textbf{c}) Small cubes, PSNR=28.2 }\\
 &&\\
 \includegraphics[width=0.3\textwidth]{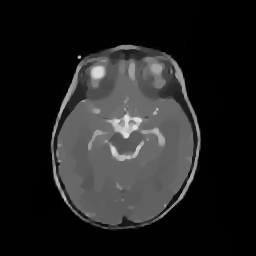}
 &
 \includegraphics[width=0.3\textwidth]{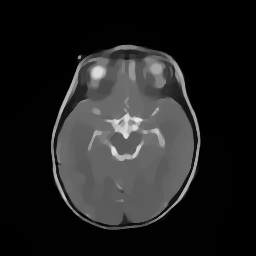} 
 & 
 \includegraphics[width=0.3\textwidth]{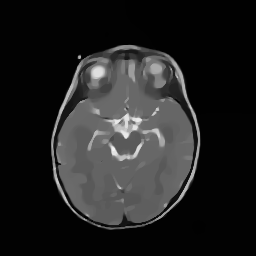} 
 \\
 \subcap{(\textbf{d}) Wavelets, PSNR = 28.1} & \subcap{(\textbf{e}) Curvelets, PSNR = 29.1} & \subcap{(\textbf{f}) Shearlets, PSNR = 30.6} 
 \end{tabular}
 \caption{Results on ``brain'' by MIND with TV regularisation and different dictionaries.}\label{f:mri}
\end{figure}

\begin{figure}
 \centering
 \begin{tabular}{ccc}
 \includegraphics[width=0.3\textwidth]{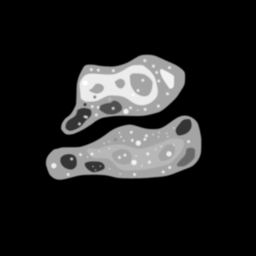} 
 &
 \includegraphics[width=0.3\textwidth]{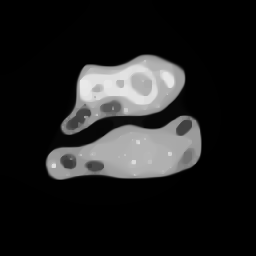}
 & 
 \includegraphics[width=0.3\textwidth]{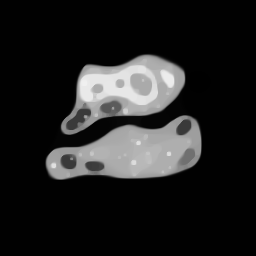}
 \\
 \subcap{(\textbf{a}) Truth} & \subcap{(\textbf{b}) Dyadic cubes, PSNR = 31.8} & \subcap{(\textbf{c}) Small cubes, PSNR = 33.6} \\
 &&\\
 \includegraphics[width=0.3\textwidth]{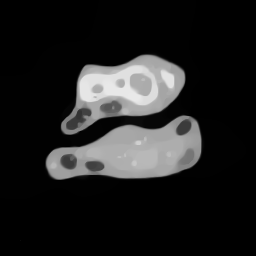}
 &
 \includegraphics[width=0.3\textwidth]{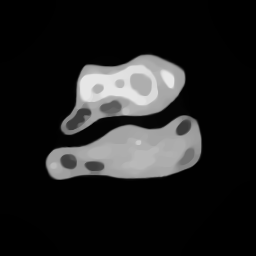}
 & 
 \includegraphics[width=0.3\textwidth]{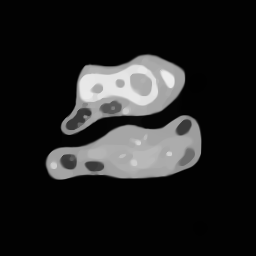}
 \\
\subcap{ (\textbf{d}) Wavelets, PSNR = 32.9} & \subcap{(\textbf{e}) Curvelets, PSNR = 32.9} & \subcap{(\textbf{f}) Shearlets, PSNR = 33.9} 
 \end{tabular}
 \caption{Results on ``cell'' by MIND with TV regularisation and different dictionaries.}\label{f:cell}
\end{figure}

\begin{figure}
 \centering
 \begin{tabular}{ccc}
 \includegraphics[width=0.3\textwidth]{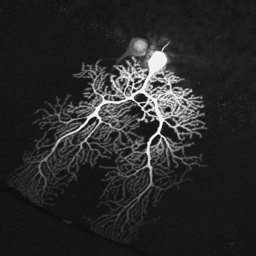} 
 &
 \includegraphics[width=0.3\textwidth]{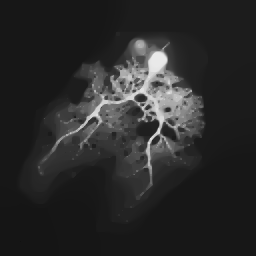} 
 & 
 \includegraphics[width=0.3\textwidth]{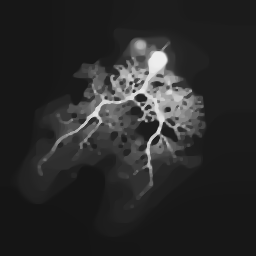} 
 \\
 \subcap{(\textbf{a}) Truth} & \subcap{(\textbf{b}) Dyadic cubes, PSNR = 25.2} & \subcap{(\textbf{c}) Small cubes, PSNR = 25.8} \\
 &&\\
 \includegraphics[width=0.3\textwidth]{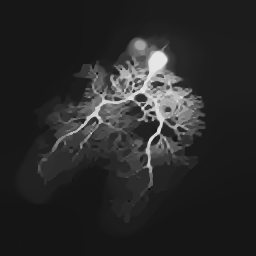}
 &
 \includegraphics[width=0.3\textwidth]{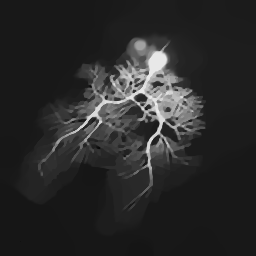} 
 & 
 \includegraphics[width=0.3\textwidth]{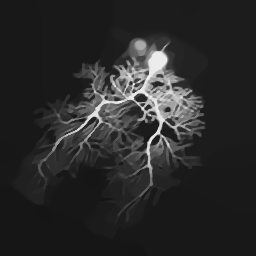} 
 \\
 \subcap{(\textbf{d}) Wavelets, PSNR = 25.8} & \subcap{(\textbf{e}) Curvelets, PSNR = 26.3} & \subcap{(\textbf{f}) Shearlets, PSNR = 27.8} 
 \end{tabular}
 \caption{Results on ``BIRN'' by MIND with TV regularisation and different dictionaries.}\label{f:birn}
\end{figure}

As a reflection of versatility of images, we consider three test images of different types. They are a magnetic resonance tomography image of mouse ``brain'  (cf. Figure \ref{f:mri}a)  from \url{Radiopaedia.org} (rID: 67777), a ``cell'' image  (cf. Figure \ref{f:cell}a)  taken from \cite{g11} and a mouse ``BIRN'' image (cf. Figure \ref{f:birn}a)  from \url{cellimagelibrary.org} (doi:10.7295/W9CCDB17) (see also  Figure \ref{f:intro} in the Introduction as well as Figure \ref{f:btfly} in Section \ref{ss:nlvl} for the case of natural images). All images are rescaled to $256 \times 256$ pixels via bicubic interpolation. The SNR is set to 20 over all cases  (see Figure \ref{f:noise} for noisy image)s. 
The results on all test images are shown in Figures \ref{f:mri}--\ref{f:birn}. In general, the dictionary of shearlets performed the best, which is followed by that of curvelets, while the rest (i.e., dyadic cubes, small cubes and wavelets) had similar performance. One exception is the ``cell'' image, where the dictionary of indicator functions of cubes, in particular of small cubes, was better at detecting tiny white balls  (see Figure \ref{f:cell}). This is because of the similarity of such features and the cubes at small scales in the dictionary. The average performance over 10 random repetitions measured by aforementioned image quality measures as well as mean integrated squared error 
$$
\mbox{MISE} = \frac{1}{n}{\sum_{i = 1}^n(\hat{f}(x_i) - f(x_i))^2}
$$
is reported in Table \ref{t:tvdict}, which is consistent with the virtual inspections in Figures \ref{f:mri}--\ref{f:birn}. {We note that 10 repetitions are sufficient here, since the variation in each repetition is comparably small (cf. standard deviations reported in parenthesis in Table \ref{t:tvdict}).} As a remark, we emphasise that the difference in performance due to the choice of dictionaries is generally negligible, but the dictionary of shearlets slightly outperforms the other choices.

\unskip

\unskip

\unskip

\begin{table}
\caption{{Average performance of MIND with $TV$ regularisation and various dictionaries, over $10$ repetitions. Values in parenthesis are standard deviations. The best value in each row is in bold.} \label{t:tvdict}}
\centering
{\scriptsize\begin{tabular}{ccccccc}
\toprule
& & \textbf{Dyadic Cubes}	& \textbf{Small Cubes} & \textbf{Wavelets} & \textbf{Curvelets} & \textbf{Shearlets}\\
\midrule
\parbox[t]{2mm}{\multirow{4}{*}{\rotatebox[origin=c]{90}{``brain''}}} 
& MISE	& 	0.00176 (5.4e-06)
		& 0.00149 (1.9e-06)	& 	0.00147 (2.9e-05)		& 0.0013 (2.5e-05) & 	\textbf{0.000871} (3e-06)		\\
& PSNR		& 	27.5 (0.013)		& 28.3 (0.0055)	& 	28.3 (0.085)		& 28.9 (0.086) & 	\textbf{30.6} (0.015)		\\
& SSIM		& 	0.871 (3.1e-06)		& 0.806 (0.002)	& 	\textbf{0.872} (0.0023)		& 0.852 (0.0091)& 	0.715 (0.0043)		\\
& VIF		& 	0.726 (0.00058)		& 0.823 (0.00036)	& 	0.767 (0.0021)		& 0.809 (0.0022)& 	\textbf{0.852} (0.00086)		\\
\midrule
\parbox[t]{2mm}{\multirow{4}{*}{\rotatebox[origin=c]{90}{``cell''}}} 
& MISE	& 	0.000671 (1.3e-06)		& 0.000438 (8.5e-07)	& 	0.000509 (5e-07)		& 0.000554 (1.3e-05) & 	\textbf{0.000412} (2.6e-06)		\\
& PSNR		& 	31.7 (0.0082)		& 33.6 (0.0084)	& 	32.9 (0.0043)		& 32.6 (0.11) & \textbf{33.9} (0.027)		\\
& SSIM		& 	0.912 (0.001)		& 0.859 (0.008)	& 	\textbf{0.924} (8.9e-05)		& 0.841 (0.018) & 	0.636 (0.0091)		\\
& VIF		& 	0.86 (8.2e-05)		& 0.913 (0.0004)	& 	0.884 (0.0004)		& 0.888 (0.00057) & \textbf{0.917} (0.00037)		\\
\midrule
\parbox[t]{2mm}{\multirow{4}{*}{\rotatebox[origin=c]{90}{``BIRN''}}} 
& MISE	& 	0.00301 (1.6e-06)		& 0.00266 (1.4e-05)	& 	0.00269 (1e-05)		& 0.00237 (7.1e-06)& 	\textbf{0.00167} (8.6e-06)		\\
& PSNR		& 	25.2 (0.0023)		& 25.7 (0.023)	& 	25.7 (0.016)		& 26.2 (0.013) & 	\textbf{27.8} (0.022)		\\
& SSIM		& 	0.791 (0.00012)		& 0.802 (0.00053)	& 	0.81 (6.5e-05)		& 0.82 (0.00075) & 	\textbf{0.858} (0.00076)		\\
& VIF		& 	0.718 (0.0018)		& 0.811 (0.0018)	& 	0.739 (0.0024)		& 0.762 (0.00096) & 	\textbf{0.838} (0.00058)		\\
\bottomrule
\end{tabular}}
\end{table}
\unskip

\subsection{{Unknown Noise Level}}\label{ss:nlvl}
\begin{figure}
 \centering
\begin{tikzpicture}[x=1pt,y=1pt]
\definecolor{fillColor}{RGB}{255,255,255}
\path[use as bounding box,fill=fillColor,fill opacity=0.00] (0,0) rectangle (252.94,180.67);
\begin{scope}
\path[clip] ( 0.00, 0.00) rectangle (252.94,180.67);
\definecolor{drawColor}{RGB}{255,255,255}
\definecolor{fillColor}{RGB}{255,255,255}

\path[draw=drawColor,line width= 0.6pt,line join=round,line cap=round,fill=fillColor] ( 0.00, 0.00) rectangle (252.94,180.68);
\end{scope}
\begin{scope}
\path[clip] ( 36.11, 30.69) rectangle (179.59,175.17);
\definecolor{fillColor}{gray}{0.92}

\path[fill=fillColor] ( 36.11, 30.69) rectangle (179.59,175.17);
\definecolor{drawColor}{RGB}{255,255,255}

\path[draw=drawColor,line width= 0.3pt,line join=round] ( 36.11, 60.79) --
	(179.59, 60.79);

\path[draw=drawColor,line width= 0.3pt,line join=round] ( 36.11,107.87) --
	(179.59,107.87);

\path[draw=drawColor,line width= 0.3pt,line join=round] ( 36.11,154.95) --
	(179.59,154.95);

\path[draw=drawColor,line width= 0.3pt,line join=round] ( 71.46, 30.69) --
	( 71.46,175.17);

\path[draw=drawColor,line width= 0.3pt,line join=round] (112.73, 30.69) --
	(112.73,175.17);

\path[draw=drawColor,line width= 0.3pt,line join=round] (154.01, 30.69) --
	(154.01,175.17);

\path[draw=drawColor,line width= 0.6pt,line join=round] ( 36.11, 37.25) --
	(179.59, 37.25);

\path[draw=drawColor,line width= 0.6pt,line join=round] ( 36.11, 84.33) --
	(179.59, 84.33);

\path[draw=drawColor,line width= 0.6pt,line join=round] ( 36.11,131.41) --
	(179.59,131.41);

\path[draw=drawColor,line width= 0.6pt,line join=round] ( 50.82, 30.69) --
	( 50.82,175.17);

\path[draw=drawColor,line width= 0.6pt,line join=round] ( 92.09, 30.69) --
	( 92.09,175.17);

\path[draw=drawColor,line width= 0.6pt,line join=round] (133.37, 30.69) --
	(133.37,175.17);

\path[draw=drawColor,line width= 0.6pt,line join=round] (174.65, 30.69) --
	(174.65,175.17);
\definecolor{fillColor}{RGB}{248,118,109}

\path[fill=fillColor,fill opacity=0.50] ( 42.63, 37.25) rectangle ( 43.94, 37.25);

\path[fill=fillColor,fill opacity=0.50] ( 43.94, 37.25) rectangle ( 45.24, 37.25);

\path[fill=fillColor,fill opacity=0.50] ( 45.24, 37.25) rectangle ( 46.55, 37.25);

\path[fill=fillColor,fill opacity=0.50] ( 46.55, 37.25) rectangle ( 47.85, 37.25);

\path[fill=fillColor,fill opacity=0.50] ( 47.85, 37.25) rectangle ( 49.15, 37.25);

\path[fill=fillColor,fill opacity=0.50] ( 49.15, 37.25) rectangle ( 50.46, 37.25);

\path[fill=fillColor,fill opacity=0.50] ( 50.46, 37.25) rectangle ( 51.76, 37.25);

\path[fill=fillColor,fill opacity=0.50] ( 51.76, 37.25) rectangle ( 53.07, 37.25);

\path[fill=fillColor,fill opacity=0.50] ( 53.07, 37.25) rectangle ( 54.37, 37.25);

\path[fill=fillColor,fill opacity=0.50] ( 54.37, 37.25) rectangle ( 55.68, 37.25);

\path[fill=fillColor,fill opacity=0.50] ( 55.68, 37.25) rectangle ( 56.98, 37.25);

\path[fill=fillColor,fill opacity=0.50] ( 56.98, 37.25) rectangle ( 58.28, 37.25);

\path[fill=fillColor,fill opacity=0.50] ( 58.28, 37.25) rectangle ( 59.59, 37.25);

\path[fill=fillColor,fill opacity=0.50] ( 59.59, 37.25) rectangle ( 60.89, 37.25);

\path[fill=fillColor,fill opacity=0.50] ( 60.89, 37.25) rectangle ( 62.20, 37.25);

\path[fill=fillColor,fill opacity=0.50] ( 62.20, 37.25) rectangle ( 63.50, 37.25);

\path[fill=fillColor,fill opacity=0.50] ( 63.50, 37.25) rectangle ( 64.81, 37.25);

\path[fill=fillColor,fill opacity=0.50] ( 64.81, 37.25) rectangle ( 66.11, 37.25);

\path[fill=fillColor,fill opacity=0.50] ( 66.11, 37.25) rectangle ( 67.41, 37.25);

\path[fill=fillColor,fill opacity=0.50] ( 67.41, 37.25) rectangle ( 68.72, 37.25);

\path[fill=fillColor,fill opacity=0.50] ( 68.72, 37.25) rectangle ( 70.02, 37.25);

\path[fill=fillColor,fill opacity=0.50] ( 70.02, 37.25) rectangle ( 71.33, 37.25);

\path[fill=fillColor,fill opacity=0.50] ( 71.33, 37.25) rectangle ( 72.63, 37.25);

\path[fill=fillColor,fill opacity=0.50] ( 72.63, 37.25) rectangle ( 73.94, 37.25);

\path[fill=fillColor,fill opacity=0.50] ( 73.94, 37.25) rectangle ( 75.24, 37.25);

\path[fill=fillColor,fill opacity=0.50] ( 75.24, 37.25) rectangle ( 76.55, 37.25);

\path[fill=fillColor,fill opacity=0.50] ( 76.55, 37.25) rectangle ( 77.85, 37.25);

\path[fill=fillColor,fill opacity=0.50] ( 77.85, 37.25) rectangle ( 79.15, 37.25);

\path[fill=fillColor,fill opacity=0.50] ( 79.15, 37.25) rectangle ( 80.46, 37.25);

\path[fill=fillColor,fill opacity=0.50] ( 80.46, 37.25) rectangle ( 81.76, 37.25);

\path[fill=fillColor,fill opacity=0.50] ( 81.76, 37.25) rectangle ( 83.07, 37.25);

\path[fill=fillColor,fill opacity=0.50] ( 83.07, 37.25) rectangle ( 84.37, 37.25);

\path[fill=fillColor,fill opacity=0.50] ( 84.37, 37.25) rectangle ( 85.68, 37.25);

\path[fill=fillColor,fill opacity=0.50] ( 85.68, 37.25) rectangle ( 86.98, 37.25);

\path[fill=fillColor,fill opacity=0.50] ( 86.98, 37.25) rectangle ( 88.28, 37.25);

\path[fill=fillColor,fill opacity=0.50] ( 88.28, 37.25) rectangle ( 89.59, 37.25);

\path[fill=fillColor,fill opacity=0.50] ( 89.59, 37.25) rectangle ( 90.89, 37.25);

\path[fill=fillColor,fill opacity=0.50] ( 90.89, 37.25) rectangle ( 92.20, 37.25);

\path[fill=fillColor,fill opacity=0.50] ( 92.20, 37.25) rectangle ( 93.50, 37.25);

\path[fill=fillColor,fill opacity=0.50] ( 93.50, 37.25) rectangle ( 94.81, 37.25);

\path[fill=fillColor,fill opacity=0.50] ( 94.81, 37.25) rectangle ( 96.11, 37.25);

\path[fill=fillColor,fill opacity=0.50] ( 96.11, 37.25) rectangle ( 97.41, 37.25);

\path[fill=fillColor,fill opacity=0.50] ( 97.41, 37.25) rectangle ( 98.72, 37.25);

\path[fill=fillColor,fill opacity=0.50] ( 98.72, 37.25) rectangle (100.02, 37.25);

\path[fill=fillColor,fill opacity=0.50] (100.02, 37.25) rectangle (101.33, 37.25);

\path[fill=fillColor,fill opacity=0.50] (101.33, 37.25) rectangle (102.63, 37.25);

\path[fill=fillColor,fill opacity=0.50] (102.63, 37.25) rectangle (103.94, 37.25);

\path[fill=fillColor,fill opacity=0.50] (103.94, 37.25) rectangle (105.24, 37.25);

\path[fill=fillColor,fill opacity=0.50] (105.24, 37.25) rectangle (106.54, 37.25);

\path[fill=fillColor,fill opacity=0.50] (106.54, 37.25) rectangle (107.85, 37.25);

\path[fill=fillColor,fill opacity=0.50] (107.85, 37.25) rectangle (109.15, 37.25);

\path[fill=fillColor,fill opacity=0.50] (109.15, 37.25) rectangle (110.46, 37.25);

\path[fill=fillColor,fill opacity=0.50] (110.46, 37.25) rectangle (111.76, 37.25);

\path[fill=fillColor,fill opacity=0.50] (111.76, 37.25) rectangle (113.07, 37.25);

\path[fill=fillColor,fill opacity=0.50] (113.07, 37.25) rectangle (114.37, 37.25);

\path[fill=fillColor,fill opacity=0.50] (114.37, 37.25) rectangle (115.67, 37.25);

\path[fill=fillColor,fill opacity=0.50] (115.67, 37.25) rectangle (116.98, 37.25);

\path[fill=fillColor,fill opacity=0.50] (116.98, 37.25) rectangle (118.28, 37.25);

\path[fill=fillColor,fill opacity=0.50] (118.28, 37.25) rectangle (119.59, 37.25);

\path[fill=fillColor,fill opacity=0.50] (119.59, 37.25) rectangle (120.89, 37.25);

\path[fill=fillColor,fill opacity=0.50] (120.89, 37.25) rectangle (122.20, 37.25);

\path[fill=fillColor,fill opacity=0.50] (122.20, 37.25) rectangle (123.50, 37.25);

\path[fill=fillColor,fill opacity=0.50] (123.50, 37.25) rectangle (124.80, 37.25);

\path[fill=fillColor,fill opacity=0.50] (124.80, 37.25) rectangle (126.11, 37.25);

\path[fill=fillColor,fill opacity=0.50] (126.11, 37.25) rectangle (127.41, 37.25);

\path[fill=fillColor,fill opacity=0.50] (127.41, 37.25) rectangle (128.72, 37.25);

\path[fill=fillColor,fill opacity=0.50] (128.72, 37.25) rectangle (130.02, 37.25);

\path[fill=fillColor,fill opacity=0.50] (130.02, 37.25) rectangle (131.33, 37.25);

\path[fill=fillColor,fill opacity=0.50] (131.33, 37.25) rectangle (132.63, 37.25);

\path[fill=fillColor,fill opacity=0.50] (132.63, 37.25) rectangle (133.93, 37.25);

\path[fill=fillColor,fill opacity=0.50] (133.93, 37.25) rectangle (135.24, 37.25);

\path[fill=fillColor,fill opacity=0.50] (135.24, 37.25) rectangle (136.54, 37.25);

\path[fill=fillColor,fill opacity=0.50] (136.54, 37.25) rectangle (137.85, 37.25);

\path[fill=fillColor,fill opacity=0.50] (137.85, 37.25) rectangle (139.15, 37.25);

\path[fill=fillColor,fill opacity=0.50] (139.15, 37.25) rectangle (140.46, 37.25);

\path[fill=fillColor,fill opacity=0.50] (140.46, 37.25) rectangle (141.76, 37.25);

\path[fill=fillColor,fill opacity=0.50] (141.76, 37.25) rectangle (143.07, 37.25);

\path[fill=fillColor,fill opacity=0.50] (143.07, 37.25) rectangle (144.37, 37.25);

\path[fill=fillColor,fill opacity=0.50] (144.37, 37.25) rectangle (145.67, 37.25);

\path[fill=fillColor,fill opacity=0.50] (145.67, 37.25) rectangle (146.98, 37.25);

\path[fill=fillColor,fill opacity=0.50] (146.98, 37.25) rectangle (148.28, 37.25);

\path[fill=fillColor,fill opacity=0.50] (148.28, 37.25) rectangle (149.59, 37.25);

\path[fill=fillColor,fill opacity=0.50] (149.59, 37.25) rectangle (150.89, 37.25);

\path[fill=fillColor,fill opacity=0.50] (150.89, 37.25) rectangle (152.20, 37.25);

\path[fill=fillColor,fill opacity=0.50] (152.20, 37.25) rectangle (153.50, 37.25);

\path[fill=fillColor,fill opacity=0.50] (153.50, 37.25) rectangle (154.80, 37.25);

\path[fill=fillColor,fill opacity=0.50] (154.80, 37.25) rectangle (156.11, 37.25);

\path[fill=fillColor,fill opacity=0.50] (156.11, 37.25) rectangle (157.41, 37.25);

\path[fill=fillColor,fill opacity=0.50] (157.41, 37.25) rectangle (158.72, 37.25);

\path[fill=fillColor,fill opacity=0.50] (158.72, 37.25) rectangle (160.02, 37.25);

\path[fill=fillColor,fill opacity=0.50] (160.02, 37.25) rectangle (161.33, 37.25);

\path[fill=fillColor,fill opacity=0.50] (161.33, 37.25) rectangle (162.63, 41.49);

\path[fill=fillColor,fill opacity=0.50] (162.63, 37.25) rectangle (163.93, 62.21);

\path[fill=fillColor,fill opacity=0.50] (163.93, 37.25) rectangle (165.24,121.53);

\path[fill=fillColor,fill opacity=0.50] (165.24, 37.25) rectangle (166.54,168.61);

\path[fill=fillColor,fill opacity=0.50] (166.54, 37.25) rectangle (167.85,164.84);

\path[fill=fillColor,fill opacity=0.50] (167.85, 37.25) rectangle (169.15,109.29);

\path[fill=fillColor,fill opacity=0.50] (169.15, 37.25) rectangle (170.46, 59.38);

\path[fill=fillColor,fill opacity=0.50] (170.46, 37.25) rectangle (171.76, 40.55);

\path[fill=fillColor,fill opacity=0.50] (171.76, 37.25) rectangle (173.06, 38.20);
\definecolor{fillColor}{RGB}{0,191,196}

\path[fill=fillColor,fill opacity=0.50] ( 42.63, 37.25) rectangle ( 43.94, 37.72);

\path[fill=fillColor,fill opacity=0.50] ( 43.94, 37.25) rectangle ( 45.24, 38.20);

\path[fill=fillColor,fill opacity=0.50] ( 45.24, 37.25) rectangle ( 46.55, 37.72);

\path[fill=fillColor,fill opacity=0.50] ( 46.55, 37.25) rectangle ( 47.85, 38.20);

\path[fill=fillColor,fill opacity=0.50] ( 47.85, 37.25) rectangle ( 49.15, 40.08);

\path[fill=fillColor,fill opacity=0.50] ( 49.15, 37.25) rectangle ( 50.46, 41.02);

\path[fill=fillColor,fill opacity=0.50] ( 50.46, 37.25) rectangle ( 51.76, 44.32);

\path[fill=fillColor,fill opacity=0.50] ( 51.76, 37.25) rectangle ( 53.07, 44.32);

\path[fill=fillColor,fill opacity=0.50] ( 53.07, 37.25) rectangle ( 54.37, 49.49);

\path[fill=fillColor,fill opacity=0.50] ( 54.37, 37.25) rectangle ( 55.68, 63.15);

\path[fill=fillColor,fill opacity=0.50] ( 55.68, 37.25) rectangle ( 56.98, 81.51);

\path[fill=fillColor,fill opacity=0.50] ( 56.98, 37.25) rectangle ( 58.28, 93.75);

\path[fill=fillColor,fill opacity=0.50] ( 58.28, 37.25) rectangle ( 59.59, 97.52);

\path[fill=fillColor,fill opacity=0.50] ( 59.59, 37.25) rectangle ( 60.89,107.40);

\path[fill=fillColor,fill opacity=0.50] ( 60.89, 37.25) rectangle ( 62.20,106.46);

\path[fill=fillColor,fill opacity=0.50] ( 62.20, 37.25) rectangle ( 63.50, 90.92);

\path[fill=fillColor,fill opacity=0.50] ( 63.50, 37.25) rectangle ( 64.81, 73.03);

\path[fill=fillColor,fill opacity=0.50] ( 64.81, 37.25) rectangle ( 66.11, 51.85);

\path[fill=fillColor,fill opacity=0.50] ( 66.11, 37.25) rectangle ( 67.41, 41.49);

\path[fill=fillColor,fill opacity=0.50] ( 67.41, 37.25) rectangle ( 68.72, 37.25);

\path[fill=fillColor,fill opacity=0.50] ( 68.72, 37.25) rectangle ( 70.02, 37.72);

\path[fill=fillColor,fill opacity=0.50] ( 70.02, 37.25) rectangle ( 71.33, 37.25);

\path[fill=fillColor,fill opacity=0.50] ( 71.33, 37.25) rectangle ( 72.63, 37.25);

\path[fill=fillColor,fill opacity=0.50] ( 72.63, 37.25) rectangle ( 73.94, 37.25);

\path[fill=fillColor,fill opacity=0.50] ( 73.94, 37.25) rectangle ( 75.24, 37.25);

\path[fill=fillColor,fill opacity=0.50] ( 75.24, 37.25) rectangle ( 76.55, 37.25);

\path[fill=fillColor,fill opacity=0.50] ( 76.55, 37.25) rectangle ( 77.85, 37.25);

\path[fill=fillColor,fill opacity=0.50] ( 77.85, 37.25) rectangle ( 79.15, 37.25);

\path[fill=fillColor,fill opacity=0.50] ( 79.15, 37.25) rectangle ( 80.46, 37.25);

\path[fill=fillColor,fill opacity=0.50] ( 80.46, 37.25) rectangle ( 81.76, 37.25);

\path[fill=fillColor,fill opacity=0.50] ( 81.76, 37.25) rectangle ( 83.07, 37.25);

\path[fill=fillColor,fill opacity=0.50] ( 83.07, 37.25) rectangle ( 84.37, 37.25);

\path[fill=fillColor,fill opacity=0.50] ( 84.37, 37.25) rectangle ( 85.68, 37.25);

\path[fill=fillColor,fill opacity=0.50] ( 85.68, 37.25) rectangle ( 86.98, 37.25);

\path[fill=fillColor,fill opacity=0.50] ( 86.98, 37.25) rectangle ( 88.28, 37.25);

\path[fill=fillColor,fill opacity=0.50] ( 88.28, 37.25) rectangle ( 89.59, 37.25);

\path[fill=fillColor,fill opacity=0.50] ( 89.59, 37.25) rectangle ( 90.89, 37.25);

\path[fill=fillColor,fill opacity=0.50] ( 90.89, 37.25) rectangle ( 92.20, 37.25);

\path[fill=fillColor,fill opacity=0.50] ( 92.20, 37.25) rectangle ( 93.50, 37.25);

\path[fill=fillColor,fill opacity=0.50] ( 93.50, 37.25) rectangle ( 94.81, 37.25);

\path[fill=fillColor,fill opacity=0.50] ( 94.81, 37.25) rectangle ( 96.11, 37.25);

\path[fill=fillColor,fill opacity=0.50] ( 96.11, 37.25) rectangle ( 97.41, 37.25);

\path[fill=fillColor,fill opacity=0.50] ( 97.41, 37.25) rectangle ( 98.72, 37.25);

\path[fill=fillColor,fill opacity=0.50] ( 98.72, 37.25) rectangle (100.02, 37.25);

\path[fill=fillColor,fill opacity=0.50] (100.02, 37.25) rectangle (101.33, 37.25);

\path[fill=fillColor,fill opacity=0.50] (101.33, 37.25) rectangle (102.63, 37.25);

\path[fill=fillColor,fill opacity=0.50] (102.63, 37.25) rectangle (103.94, 37.25);

\path[fill=fillColor,fill opacity=0.50] (103.94, 37.25) rectangle (105.24, 37.25);

\path[fill=fillColor,fill opacity=0.50] (105.24, 37.25) rectangle (106.54, 37.25);

\path[fill=fillColor,fill opacity=0.50] (106.54, 37.25) rectangle (107.85, 37.25);

\path[fill=fillColor,fill opacity=0.50] (107.85, 37.25) rectangle (109.15, 37.25);

\path[fill=fillColor,fill opacity=0.50] (109.15, 37.25) rectangle (110.46, 37.25);

\path[fill=fillColor,fill opacity=0.50] (110.46, 37.25) rectangle (111.76, 37.25);

\path[fill=fillColor,fill opacity=0.50] (111.76, 37.25) rectangle (113.07, 37.25);

\path[fill=fillColor,fill opacity=0.50] (113.07, 37.25) rectangle (114.37, 37.25);

\path[fill=fillColor,fill opacity=0.50] (114.37, 37.25) rectangle (115.67, 37.25);

\path[fill=fillColor,fill opacity=0.50] (115.67, 37.25) rectangle (116.98, 37.25);

\path[fill=fillColor,fill opacity=0.50] (116.98, 37.25) rectangle (118.28, 37.25);

\path[fill=fillColor,fill opacity=0.50] (118.28, 37.25) rectangle (119.59, 37.25);

\path[fill=fillColor,fill opacity=0.50] (119.59, 37.25) rectangle (120.89, 37.25);

\path[fill=fillColor,fill opacity=0.50] (120.89, 37.25) rectangle (122.20, 37.25);

\path[fill=fillColor,fill opacity=0.50] (122.20, 37.25) rectangle (123.50, 37.25);

\path[fill=fillColor,fill opacity=0.50] (123.50, 37.25) rectangle (124.80, 37.25);

\path[fill=fillColor,fill opacity=0.50] (124.80, 37.25) rectangle (126.11, 37.25);

\path[fill=fillColor,fill opacity=0.50] (126.11, 37.25) rectangle (127.41, 37.25);

\path[fill=fillColor,fill opacity=0.50] (127.41, 37.25) rectangle (128.72, 37.25);

\path[fill=fillColor,fill opacity=0.50] (128.72, 37.25) rectangle (130.02, 37.25);

\path[fill=fillColor,fill opacity=0.50] (130.02, 37.25) rectangle (131.33, 37.25);

\path[fill=fillColor,fill opacity=0.50] (131.33, 37.25) rectangle (132.63, 37.25);

\path[fill=fillColor,fill opacity=0.50] (132.63, 37.25) rectangle (133.93, 37.25);

\path[fill=fillColor,fill opacity=0.50] (133.93, 37.25) rectangle (135.24, 37.25);

\path[fill=fillColor,fill opacity=0.50] (135.24, 37.25) rectangle (136.54, 37.25);

\path[fill=fillColor,fill opacity=0.50] (136.54, 37.25) rectangle (137.85, 37.25);

\path[fill=fillColor,fill opacity=0.50] (137.85, 37.25) rectangle (139.15, 37.25);

\path[fill=fillColor,fill opacity=0.50] (139.15, 37.25) rectangle (140.46, 37.25);

\path[fill=fillColor,fill opacity=0.50] (140.46, 37.25) rectangle (141.76, 37.25);

\path[fill=fillColor,fill opacity=0.50] (141.76, 37.25) rectangle (143.07, 37.25);

\path[fill=fillColor,fill opacity=0.50] (143.07, 37.25) rectangle (144.37, 37.25);

\path[fill=fillColor,fill opacity=0.50] (144.37, 37.25) rectangle (145.67, 37.25);

\path[fill=fillColor,fill opacity=0.50] (145.67, 37.25) rectangle (146.98, 37.25);

\path[fill=fillColor,fill opacity=0.50] (146.98, 37.25) rectangle (148.28, 37.25);

\path[fill=fillColor,fill opacity=0.50] (148.28, 37.25) rectangle (149.59, 37.25);

\path[fill=fillColor,fill opacity=0.50] (149.59, 37.25) rectangle (150.89, 37.25);

\path[fill=fillColor,fill opacity=0.50] (150.89, 37.25) rectangle (152.20, 37.25);

\path[fill=fillColor,fill opacity=0.50] (152.20, 37.25) rectangle (153.50, 37.25);

\path[fill=fillColor,fill opacity=0.50] (153.50, 37.25) rectangle (154.80, 37.25);

\path[fill=fillColor,fill opacity=0.50] (154.80, 37.25) rectangle (156.11, 37.25);

\path[fill=fillColor,fill opacity=0.50] (156.11, 37.25) rectangle (157.41, 37.25);

\path[fill=fillColor,fill opacity=0.50] (157.41, 37.25) rectangle (158.72, 37.25);

\path[fill=fillColor,fill opacity=0.50] (158.72, 37.25) rectangle (160.02, 37.25);

\path[fill=fillColor,fill opacity=0.50] (160.02, 37.25) rectangle (161.33, 37.25);

\path[fill=fillColor,fill opacity=0.50] (161.33, 37.25) rectangle (162.63, 37.25);

\path[fill=fillColor,fill opacity=0.50] (162.63, 37.25) rectangle (163.93, 37.25);

\path[fill=fillColor,fill opacity=0.50] (163.93, 37.25) rectangle (165.24, 37.25);

\path[fill=fillColor,fill opacity=0.50] (165.24, 37.25) rectangle (166.54, 37.25);

\path[fill=fillColor,fill opacity=0.50] (166.54, 37.25) rectangle (167.85, 37.25);

\path[fill=fillColor,fill opacity=0.50] (167.85, 37.25) rectangle (169.15, 37.25);

\path[fill=fillColor,fill opacity=0.50] (169.15, 37.25) rectangle (170.46, 37.25);

\path[fill=fillColor,fill opacity=0.50] (170.46, 37.25) rectangle (171.76, 37.25);

\path[fill=fillColor,fill opacity=0.50] (171.76, 37.25) rectangle (173.06, 37.25);
\definecolor{drawColor}{RGB}{255,0,0}

\path[draw=drawColor,line width= 0.6pt,line join=round] ( 75.58, 30.69) -- ( 75.58,175.17);
\end{scope}
\begin{scope}
\path[clip] ( 0.00, 0.00) rectangle (252.94,180.67);
\definecolor{drawColor}{gray}{0.30}

\node[text=drawColor,anchor=base east,inner sep=0pt, outer sep=0pt, scale= 0.68] at ( 31.16, 34.22) {0};

\node[text=drawColor,anchor=base east,inner sep=0pt, outer sep=0pt, scale= 0.68] at ( 31.16, 81.30) {100};

\node[text=drawColor,anchor=base east,inner sep=0pt, outer sep=0pt, scale= 0.68] at ( 31.16,128.38) {200};
\end{scope}
\begin{scope}
\path[clip] ( 0.00, 0.00) rectangle (252.94,180.67);
\definecolor{drawColor}{gray}{0.20}

\path[draw=drawColor,line width= 0.6pt,line join=round] ( 33.36, 37.25) --
	( 36.11, 37.25);

\path[draw=drawColor,line width= 0.6pt,line join=round] ( 33.36, 84.33) --
	( 36.11, 84.33);

\path[draw=drawColor,line width= 0.6pt,line join=round] ( 33.36,131.41) --
	( 36.11,131.41);
\end{scope}
\begin{scope}
\path[clip] ( 0.00, 0.00) rectangle (252.94,180.67);
\definecolor{drawColor}{gray}{0.20}

\path[draw=drawColor,line width= 0.6pt,line join=round] ( 50.82, 27.94) --
	( 50.82, 30.69);

\path[draw=drawColor,line width= 0.6pt,line join=round] ( 92.09, 27.94) --
	( 92.09, 30.69);

\path[draw=drawColor,line width= 0.6pt,line join=round] (133.37, 27.94) --
	(133.37, 30.69);

\path[draw=drawColor,line width= 0.6pt,line join=round] (174.65, 27.94) --
	(174.65, 30.69);
\end{scope}
\begin{scope}
\path[clip] ( 0.00, 0.00) rectangle (252.94,180.67);
\definecolor{drawColor}{gray}{0.30}

\node[text=drawColor,anchor=base,inner sep=0pt, outer sep=0pt, scale= 0.68] at ( 50.82, 19.68) {12};

\node[text=drawColor,anchor=base,inner sep=0pt, outer sep=0pt, scale= 0.68] at ( 92.09, 19.68) {13};

\node[text=drawColor,anchor=base,inner sep=0pt, outer sep=0pt, scale= 0.68] at (133.37, 19.68) {14};

\node[text=drawColor,anchor=base,inner sep=0pt, outer sep=0pt, scale= 0.68] at (174.65, 19.68) {15};
\end{scope}
\begin{scope}
\path[clip] ( 0.00, 0.00) rectangle (252.94,180.67);
\definecolor{drawColor}{RGB}{0,0,0}

\node[text=drawColor,anchor=base,inner sep=0pt, outer sep=0pt, scale= 0.9] at (107.85, 7.64) {Noise level};
\end{scope}
\begin{scope}
\path[clip] ( 0.00, 0.00) rectangle (252.94,180.67);
\definecolor{drawColor}{RGB}{0,0,0}

\node[text=drawColor,rotate= 90.00,anchor=base,inner sep=0pt, outer sep=0pt, scale= 0.9] at ( 13.08,102.93) {Count};
\end{scope}
\begin{scope}
\path[clip] ( 0.00, 0.00) rectangle (252.94,180.67);
\definecolor{fillColor}{RGB}{255,255,255}

\path[fill=fillColor] (190.59, 80.23) rectangle (247.44,125.63);
\end{scope}
\begin{scope}
\path[clip] ( 0.00, 0.00) rectangle (252.94,180.67);
\definecolor{fillColor}{gray}{0.95}

\path[fill=fillColor] (196.09,100.18) rectangle (210.54,114.63);
\end{scope}
\begin{scope}
\path[clip] ( 0.00, 0.00) rectangle (252.94,180.67);
\definecolor{fillColor}{RGB}{248,118,109}

\path[fill=fillColor,fill opacity=0.50] (196.80,100.89) rectangle (209.83,113.92);
\end{scope}
\begin{scope}
\path[clip] ( 0.00, 0.00) rectangle (252.94,180.67);
\definecolor{fillColor}{gray}{0.95}

\path[fill=fillColor] (196.09, 85.73) rectangle (210.54,100.18);
\end{scope}
\begin{scope}
\path[clip] ( 0.00, 0.00) rectangle (252.94,180.67);
\definecolor{fillColor}{RGB}{0,191,196}

\path[fill=fillColor,fill opacity=0.50] (196.80, 86.44) rectangle (209.83, 99.47);
\end{scope}
\begin{scope}
\path[clip] ( 0.00, 0.00) rectangle (252.94,180.67);
\definecolor{drawColor}{RGB}{0,0,0}

\node[text=drawColor,anchor=base west,inner sep=0pt, outer sep=0pt, scale= 0.88] at (216.04,104.38) {by \cite{munk2005difference}};
\end{scope}
\begin{scope}
\path[clip] ( 0.00, 0.00) rectangle (252.94,180.67);
\definecolor{drawColor}{RGB}{0,0,0}

\node[text=drawColor,anchor=base west,inner sep=0pt, outer sep=0pt, scale= 0.88] at (216.04, 89.92) {by \cite{LTO13}};
\end{scope}
\end{tikzpicture}
\caption{{Performance of two noise level estimators by \citet{munk2005difference} and   \citet{LTO13} over 1000 repetitions, with the true $\sigma$ indicated by a vertical red line, in case of ``butterfly'' image with SNR = 20.}}
\label{f:estNL}
\end{figure}
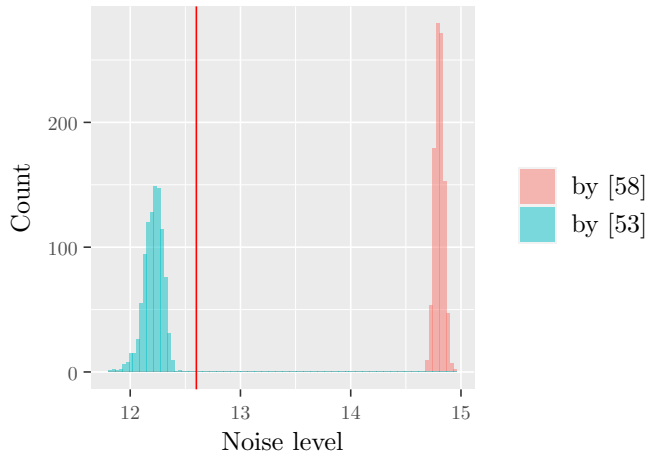

\begin{figure}
\centering
\begin{tabular}{cc}
 \includegraphics[width=0.3\textwidth]{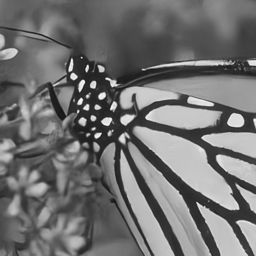} 
 &
 \includegraphics[width=0.3\textwidth]{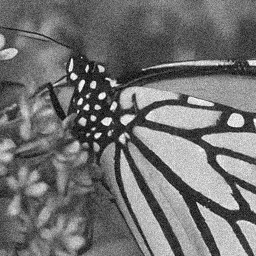} 
 \\
 \subcap{(\textbf{a}) Truth} & \subcap{(\textbf{b}) Noisy image, PSNR = 26.1} \\
 &
 \end{tabular}
{\begin{tabular}{ccc}
 \includegraphics[width=0.3\textwidth]{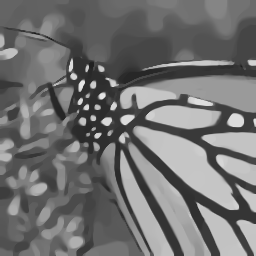}
 &
 \includegraphics[width=0.3\textwidth]{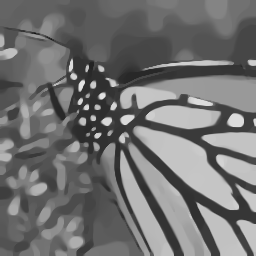} 
 & 
 \includegraphics[width=0.3\textwidth]{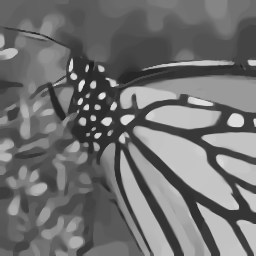} 
 \\
 \subcap{(\textbf{c}) MIND with true $\sigma$, PSNR=28.9} & \subcap{(\textbf{d}) MIND with $\hat\sigma$ by \cite{munk2005difference}, PSNR = 28} & \subcap{(\textbf{e}) MIND with $\hat\sigma$ by \cite{LTO13}, PSNR = 29.1} 
 \end{tabular}}
 \caption{Results on ``butterfly'' with unknown noise level: (\textbf{c}--\textbf{e}) MIND with total variation and shearlets dictionary is shown, with threshold $q_n$ the 50\%-quantile of $\|\tilde\sigma\epsilon\|_{\text{MS}}$, where $\tilde \sigma$ is true $\sigma$, estimated $\hat \sigma$ by \citet{munk2005difference} and   \citet{LTO13}, respectively.}\label{f:btfly}
\end{figure}

{
We now investigate the issue of unknown noise level $\sigma$. This   is illustrated by implementing two different estimators for the noise level into the MIND estimator with total variation regularisation and shearlets dictionary. We stress, however, that the results are to be expected to be similar for other choices of the dictionary and regularisation functional. The MIND optimisation is solved by the Chambolle--Pock algorithm. However, in contrast to previous simulations the 50\%-quantile of $\|\sigma\epsilon\|_{\text{MS}}$ in \eqref{Residual} with known $\sigma$ is now replaced by an estimated noise level $\hat\sigma$ which then gives a new threshold $q_n$ for MIND. We display all results for the well known test image ``butterfly'' (\mbox{$256 \times 256$ pixels}, see,   e.g., \cite{ZhEt17}). The SNR is set to 20, or equivalently $\sigma=12.6$. The choice of test image and SNR is purely arbitrary,   and   the results would remain quite the same for other choices. We investigate the influence on MIND for two different noise level estimators. One is a second-order difference-based estimator by \citet{munk2005difference}, which tends to overestimate the noise level,   and   the other one is a patch based estimator by \citet{LTO13}, which often underestimates the noise level  (see Figure \ref{f:estNL}). The comparison results are summarised in Figure \ref{f:btfly}. Visually, there is little difference between the use of MIND with known noise level and that of the estimated ones. A slight improvement in terms of PSNR is even shown for the use of the estimator that tends to underestimate the noise level. However, a caution should be taken that an underestimated noise level leads to a sacrifice in statistical confidence. For instance, it is no longer  possible   to guarantee that the true image $f$ lies in the constraint of \eqref{MIND} with given probability 50\%. In contrast, when using the true noise level or an overestimated noise level, such a statistical confidence statement is valid. In short, there is nearly no loss of performance of MIND due to the unknown noise level in practice. A heuristic explanation of this is that the mean squared error of the estimated noise level $\sigma$ scaled as $O(n^{-1})$, whereas for estimating $f$ this rate is slower. Hence, its effect on the overall performance is of minor order. 
}

\section{{Conclusions and Discussion}}\label{CD}

\subsection{{Conclusions}}
In this paper, we   discuss  three different algorithms for the computation of variational multiscale estimators (MIND). Among them, we    find  the Chambolle--Pock algorithm to perform best in practice and to be equally suited for smooth and non-smooth regularisation functionals $R(\cdot)$, the latter case including the prominent TV regularisation functional. Further, there is no need to tune parameters that are related to the Chambolle--Pock algorithm, as the default choice already delivers desirable performance.

Concerning the MIND estimator itself, we   find that, for two-dimensional imaging, in nearly all cases it performs best when employing shearlets as the dictionary. However, if more specific a-priori knowledge about the shape of features is available, improvements can be achieved with a dictionary consisting of basis functions that resemble such features. The MIND estimator can be easily modified accordingly. For example, as we   demonstrate, in the  case of images with bubble-like structures, the choice of cubes as dictionary leads to very promising results.

In contrast to many other statistical recovery methods relying on subtle regularisation parameter choices which are often   difficult to choose and interpret, the only relevant parameter for MIND (besides parameters for the numerical optimisation step) is a significance level $\alpha$ between 0 and 1. This has a clear statistical meaning: it guarantees that the estimator is no rougher than the truth in terms of the regularisation functional $R(\cdot)$ with probability at least $1-\alpha$.

{Finally, the unknown noise level can be easily estimated by a difference-based or patch-based estimator, both being  easy to compute in $O(n)$ steps.}

\subsection{{Extensions}}
\subsubsection{{Bump signals and inverse problems}}

Even though our findings give a quite satisfactory answer to the question of computability of variational multiscale estimators, some additional questions arise. The first one concerns model \eqref{model}. We   focus  on the nonparametric regression model,   and   in particular on image denoising. Concerning this, we stress that variational multiscale estimators can be applied to other settings, as well. The most obvious one is signal recovery when $d=1$,   and   specific structure of signals can be incorporated as well, such as locally constant signals \cite{Frick2014}, where the number of constant segments is incorporated into the regularisation functional. Although the resulting functional is no longer convex,   the MIND estimator can   still be computed efficiently exploiting a dynamic program.

A further extensions concerns general linear inverse problems in any dimension $d$ (see, e.g., \cite{del2019total}). In such noisy inverse problems, the only difference concerns the dictionary $\{\phi_{\lambda}\}$, which has to be chosen in a way akin to the wavelet-vaguelette transform \cite{donoho1995nonlinear}. Otherwise,   the structure of the estimator \eqref{MIND} remains unchanged,   and   in particular the optimisation problem to be solved remains the same. We hence expect the findings of this paper to apply to variational multiscale estimators for inverse problems as well.

\subsubsection{{Different Noise Models}}

The second possible extension of the present paper concerns the noise model. We consider Gaussian noise with homogeneous (i.e., not depending on the spatial location $x$) yet unknown variance. We remark that the extension to heterogeneous variance is of interest as in many applications the variance varies with the signal. To this end, the residuals $Y-f$ in (\ref{MIND}) have to be standardised by a local estimator of the variance  (see,   e.g., \cite{brown2007}).

We finally mention that variational multiscale estimators have been proposed for non-Gaussian noise models as well (see,   e.g., \cite{fmm13} for Poisson noise). For non-Gaussian data, the constraint \mbox{$|\langle \phi_{\lambda},g-Y\rangle|\leq q_n$} in \eqref{MIND} is replaced by a constraint on the likelihood-ratio statistic, which offers a route to generalise this further, e.g., to exponential family models  (see \cite{Koenig2020}). In some cases (e.g., Poisson), after a variance stabilising transformation, that constraint can be turned into a linear constraint as in the Gaussian case, so our findings are expected to apply there as well. For general noise, however, deeper study will be required, as nonlinearity in the constraint may lead to nonconvexity of the optimisation problem \eqref{MIND}.

\section*{Acknowledgement}
MA  is  supported  by the Deutsche Forschungsgemeinschaft  (DFG; German Research Foundation) Postdoctoral Fellowship AL 2483/1-1. HL and AM are funded by the Deutsche Forschungsgemeinschaft (DFG, German Research Foundation) under Germany’s Excellence Strategy - EXC 2067/1-390729940. The authors  would like to thank Timo Aspelmeier for providing code for the ADMM based implementation.


\end{document}